\documentclass[10pt, conference]{IEEEtran}

\usepackage{amsmath,amsfonts}
\usepackage{amssymb}
\usepackage{algorithm}
\usepackage[noend]{algpseudocode} 
\usepackage{etoolbox}

\newcommand{\rem}[1]{}

\usepackage{graphicx}
\usepackage{textcomp}
\usepackage{xcolor}
\usepackage{verbatim}

\usepackage{lineno}
\usepackage{listings}
\usepackage{url}
\usepackage{hyperref}
\usepackage{cleveref}
\usepackage{verbatim}
\usepackage{enumerate}
\usepackage{graphicx}
\usepackage{subcaption}
\usepackage{booktabs}
\usepackage{multicol}
\usepackage{multirow}
\usepackage{tabularx}
\usepackage{enumitem}
\usepackage{cite}
\usepackage{wrapfig}

\providecommand{\linesref}[2]{\hyperref[#1]{Lines~\ref*{#1}--\ref*{#2}}}
\providecommand{\Linesref}[2]{\hyperref[#1]{Lines~\ref*{#1}} \hyperref[#2]{and~\ref*{#2}}}
\providecommand{\lineref}[1]{\hyperref[#1]{Line~\ref*{#1}}}
\providecommand{\Lineref}[1]{\hyperref[#1]{Line~\ref*{#1}}}

\def\BibTeX{{\rm B\kern-.05em{\sc i\kern-.025em b}\kern-.08em
    T\kern-.1667em\lower.7ex\hbox{E}\kern-.125emX}}

\usepackage{xpatch}
\makeatletter
\xpatchcmd{\algorithmic}{\itemsep\z@}{\itemsep=0.5ex plus1pt minus0.4ex}{}{}
\makeatother

\begin{document}


\title{Parallel Algorithms and Heuristics for Efficient Computation of High-Order Line Graphs of Hypergraphs \vspace{-1em}}



\author{
  \IEEEauthorblockN{
  Xu T. Liu \IEEEauthorrefmark{1}\IEEEauthorrefmark{2},
  Jesun Firoz\IEEEauthorrefmark{2},
  Andrew Lumsdaine\IEEEauthorrefmark{2}\IEEEauthorrefmark{3},
  Cliff Joslyn\IEEEauthorrefmark{2},\\
  Sinan Aksoy\IEEEauthorrefmark{2},
  Brenda Praggastis\IEEEauthorrefmark{2},
  Assefaw H. Gebremedhin\IEEEauthorrefmark{1}
  }
  \IEEEauthorblockA{
    \IEEEauthorrefmark{1}Washington State University, 
    \IEEEauthorrefmark{2}Pacific Northwest National Lab, 
    \IEEEauthorrefmark{3}University of Washington
  }
  \IEEEauthorblockA{
    \IEEEauthorrefmark{1}\{xu.liu2, assefaw.gebremedhin\}@wsu.edu,
    \IEEEauthorrefmark{2}\{jesun.firoz, andrew.lumsdaine, cliff.joslyn, sinan.aksoy, brenda.praggastis\}@pnnl.gov
  }
}

\maketitle
\thispagestyle{plain}
\pagestyle{plain}

\begin{abstract}
This paper considers structures of systems beyond dyadic (pairwise) interactions and investigates mathematical modeling of multi-way interactions and connections as hypergraphs, where captured relationships among system entities are set-valued. To date, in most situations, entities in a hypergraph are considered connected as long as there is at least one common ``neighbor''. However, minimal commonality sometimes discards the ``strength'' of connections and interactions among groups. To this end, considering the ``width'' of a connection, referred to as the \emph{$s$-overlap} of neighbors, provides more meaningful insights into how closely the communities or entities interact with each other. In addition, $s$-overlap computation is the fundamental kernel to construct the line graph of a hypergraph, a low-order approximation of the hypergraph which can carry significant information about the original hypergraph. Subsequent stages of a data analytics pipeline then can apply highly-tuned graph algorithms on the line graph to reveal important features. Given a hypergraph, computing the $s$-overlaps by exhaustively considering all pairwise entities can be computationally prohibitive. 
To tackle this challenge, we develop efficient algorithms to compute $s$-overlaps and the corresponding line graph of a hypergraph. We propose several heuristics to avoid execution of redundant work and improve performance of the $s$-overlap computation. {\color{black}Our parallel algorithm, combined with these heuristics, 
is orders of magnitude (more than $10\times$) faster than the naive algorithm in all cases and the SpGEMM algorithm with filtration in most cases (especially with large $s$ value).} 
\end{abstract}

\begin{IEEEkeywords}
Hypergraphs, parallel hypergraph algorithms, line graphs, intersection graphs.
\end{IEEEkeywords}

\section{Introduction}\label{sec:introduction}
 
Graph-theoretical mathematical abstractions 
represent entities of interest as \emph{vertices} connected by  \emph{edges} between pairs of them. 
Classical graph models benefit from simplicity and a degree of universality. But as abstract mathematical objects, graphs are limited to representing {\it pairwise} relationships between entities, whereas real-world phenomena in these systems can be rich in {\em multi-way} relationships involving interactions among more than two entities, dependencies between more than two variables, or properties of collections of more than two objects. 
Representing {\em group} interactions is not possible in graphs natively, but rather requires either more complex mathematical objects, or coding schemes like ``reification'' or  semantic labeling in bipartite graphs.
Lacking multi-dimensional relations, it is hard to address questions of ``community interaction'' in graphs: e.g.,\ how is a collection of
entities $A$ connected to another collection $B$ through chains of
other communities; where does a particular community stand in
relation to other communities in its neighborhood.

\begin{figure}[t]
     \centering
    \begin{subfigure}[t]{.45\linewidth}
       \centering\includegraphics[width=\linewidth]{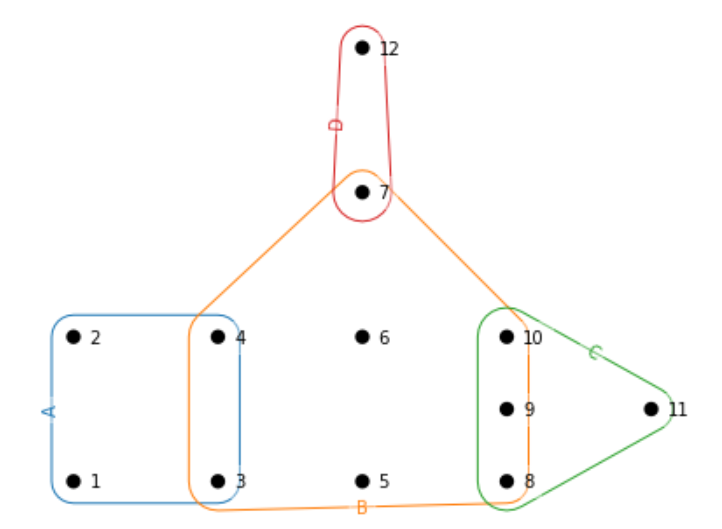}
        \caption{\small \textit{ A hypergraph, $H$.}} \label{fig:hypergraph_example}
    \end{subfigure}
~
    \begin{subfigure}[t]{.45\linewidth}
       \centering\includegraphics[width=\linewidth]{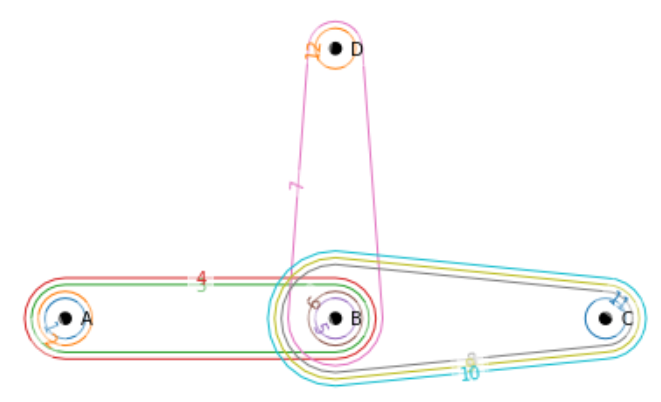}
        \caption{\small \textit{ The dual hypergraph of $H$, $H^{*}$.} }
    \end{subfigure}
    \begin{subfigure}[t]{\linewidth}
        \centering\includegraphics[width=0.9\linewidth]{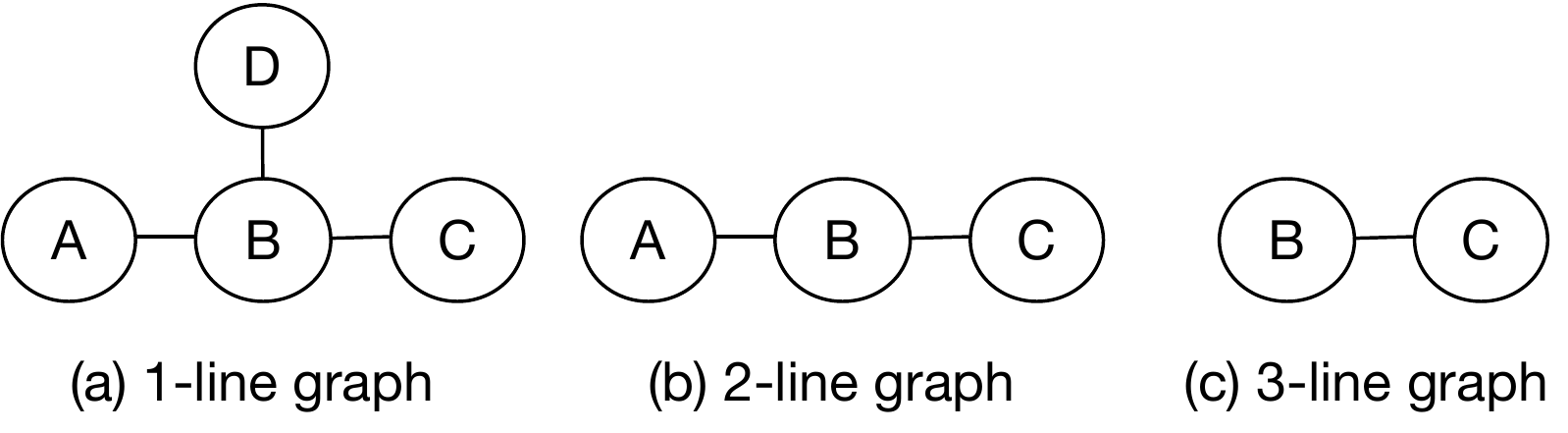}
        \caption{\small \textit {The line graphs $L_s(H)$ of hypergraph $H$ for $s=1,2,3$.} }\label{fig:linegraph}
    \end{subfigure}    
\vspace{1em}    
  \caption{\small \textit{Euler diagram of a hypergraph $H$ and its dual $H^{*}$. $H$ has hyperedges  $ E=\{ A, B, C, D \}$ on the vertex set  $V=\{1, 2, \ldots, 12 \}$. Dually, $H^*$ has edge set $E^* = \{1, 2, \ldots, 12 \}$ with vertex set $V^*= \{ A, B, C, D \}.$
\rem{Edge $Blue$ contains vertices $\{A,B,C,D,E,L,M\}$. Similarly, the vertex set for the hyperedge $Orange:\{L,E,M,F,G,O,N,H,P\}$ and hyperedge $Green:\{O,N,H,P,I,J\}$.}
  The diagram shows each of the four hyperedges as a ``lasso'' around its vertices. The visualization was done with the HyperNetX library~\cite{praggastispnnl}. (c) The $s$-line graphs for $H$ for $s=1,2,3$ } } 
  \label{fig:euler_diagram_hypergraph}
\end{figure}

\rem{However, since graph edges only relate pairs of vertices, a graph-theoretical abstraction will fail to natively capture multiway relationships among entities~\cite{aksoy2020hypernetwork}.  }

The mathematical object that {\em natively} represents multi-way interactions in networks is called a  ``hypergraph''  \cite{berge1973graphs}. %
In contrast to a graph, in a hypergraph $H = (V, E)$ those same vertices are now connected generally in a family $E$ of hyperedges, where now a hyperedge $e \in E$ is an arbitrary subset $e \subseteq V$ of $k$ vertices, thereby representing a $k$-way relationship for any integer $k > 0$.
Hypergraphs are thus the natural representation of a broad range of systems, including those with the kinds of multi-way relationships mentioned above.
Indeed, hypergraph-structured data (i.e.\ hypernetworks) are ubiquitous, occurring whenever information presents naturally as set-valued, tabular, or bipartite data. 
Since hyperedge incidence and vertex adjacency are set-valued (\Cref{fig:euler_diagram_hypergraph}), we use hyperedge and vertex interchangeably.
 
\rem{
\textit{Hypergraphs} 
 generalize graphs by defining \emph{hyperedges}  between any number of vertices instead of only two, thus being able to represent multi-way relationships naturally. }

An example of a hypergraph $H=(V,E)$ is shown in (\Cref{fig:euler_diagram_hypergraph}). Here the vertex set is  $V=\{1, 2, \ldots, 12 \}$, and there are four hyperedges $e \in E = \{A,B,C,D\}$, where each $e \subseteq V$. For example, $B = \{3,4,\ldots,10\} \in E$. Note that edges vary in size, with e.g.\ $|A|=4,|B|=8$. Note also that $|D|=2$. Thus $D$, and $D$ alone, is a proper graph edge. In fact, every graph is a 2-uniform hypergraph, where every edge has size 2. Every hypergraph $H$ has a dual $H^*$ constructed by considering the set $E$ as a new vertex set $V^{*}$ and the set $V$ as hyperedge set, $E^{*}$ in $H^{*}$.  In graphs, incidence edges share a common vertex, while in hypergraph, incident hyperedges can vary in intersection size. For example, $|A \cap B| = 2$, while $|B \cap C|=3$ and $|B \cap D|= 1$. We use the variable $s$ to indicate edge intersection size, and call that intersection the \emph{$s$-overlap} between those edges.

  \rem{Note that, $(Orange,Green)$ edge pair has a maximum $s$-overlap of 4 and $(Blue, Green)$ edge pair has a $s$-overlap of at most 3, depicting the maximum width possible for each such connected pairs of hyperedges. Since the incidence (neighboring vertices) for each edge and the adjacency for each vertex are set-valued. Hence the dual hypergraph $H^{*}$ can be  }

\rem{For example, in biomolecular modeling, to explore the relationships among human genes and diseases, in particular, to better understand potential overlaps between human gene sets and  metabolic  rare  diseases, and their known biological processes and chemical interactions, hypergraph metrics have been applied~\cite{FeSHeE20,joslyn2020hypernetwork}. In addition, hypergraph analytics has been shown~\cite{joslyn2020hypergraph} to be very useful to model complex multiway relations in cyber data, i.e. among domains and IP addresses.} 


 
 \rem{, hypergraph-related algorithms, more specifically, algorithms based on hypergraph walks (i.e. algorithms for hypergraph traversal and connectivity) mostly  considered \textit{connectivity width} of 1, where, as long as there is at least one \textit{overlapping} vertex that belongs to the neighbor sets of two  hyperedge endpoints, the hyperedges are considered walkable from one to the other.}


       

Computation of $s$-overlaps of a hypergraph to find all pairs of connecting hyperedges with minimal $s$ common neighbors is the most fundamental kernel in the general framework of hypergraph analytics for connectivity and traversal~\cite{aksoy2020hypernetwork}. 
As we will show, hypergraph structures naturally stratify into sub-structures parameterized by $s$-overlap, each of which lends itself to being exploited by traditional, highly efficient graph algorithms. In particular, the $s$-line graph of a hypergraph is a form of dual graph (not a hypergraph) now on the set of hyperedges as vertices, where each pair of hyperedges are connected if they have an $s$-overlap. \Cref{fig:linegraph} shows the $s$-line graphs for $s=1,2,3$ for $H$ from \Cref{fig:hypergraph_example}.

\begin{table}[t]
\centering
\begin{tabular}{l|rr}
\hline
Stage                 & Naive             & Our method       \\ \hline
\textbf{$s$-overlap}     & \textbf{314.757s} & \textbf{0.684s} \\
Squeeze               & 0.005s            & 0.005s          \\
$s$-connected component & \textless{}1ms    & \textless{}1ms  \\ \hline
Total time            & 314.762s          & 0.689s          \\
Speedup               & 1$\times$         & 460$\times$     \\
\#set intersections    & $3.52 \times 10^{10}$ & $3.26 \times 10^{7}$       \\ \hline
\end{tabular}
\caption{\small \color{black} \textit{Computational cost of each step of the pipeline to compute the $s$-line graph from the email-EuAll dataset~\cite{suitsparsecollection}. Clearly, $s$-overlap computation (in bold) time is the dominant stage in the process.}}\label{tab:comp_cost}
\vspace{-1.5em}
\end{table}

{\color{black}
There are three main motivations for computing the $s$-line graph of a hypergraph. First, optimized graph kernels can be computed on the $s$-line graph directly to compute important metrics such as $s$-connected components, $s$-centralities etc. Second, $s$-line graph computation enables structural analyses of hypergraphs that are otherwise challenging or impossible to determine without their formation.
As we illustrate in \Cref{sec:motivating}, one such example is spectral analyses of hypergraphs, which are based on eigenvalues and eigenvectors of $s$-line graph matrices. In general, $s$-line graph matrices must be formed explicitly in order for their entire spectrum to be computed, and cannot be inferred from the incidence matrix representing the hypergraph. 
Third, the application of $s$-line graph can be found in many important domains~\cite{aksoy2020hypernetwork,FeSHeE20}.
For example, in~\cite{FeSHeE20},  $s$-overlap walks are used to define hypergraph $s$-centralities applied to gene expression levels in virology data sets. Here vertices represent experimental conditions and  hyperedges represent genes. They were able to show a much higher level of gene enhancement measurement scores when taking high-order $s$-overlaps into account (including $s$ values well above 10).
}

Computing the $s$-line graph of a hypergraph on large datasets can become computationally challenging when considering each possible combinations of pair of hyperedges (~\Cref{tab:comp_cost}). 
To the best of our knowledge, the only-available, sequential implementation of the $s$-overlap computation can be found in the python-based HyperNetX~\cite{praggastispnnl} library with NetworkX~\cite{hagberg2008exploring} as the backend. However, the algorithm implemented in HyperNetX ( 
naive algorithm discussed in \Cref{sec:our_algorithm}) is quite inefficient and fails to execute with larger datasets that we consider in~\Cref{sec:evaluation}. 
Hence devising efficient, parallel algorithm to compute $s$-line graph is vital. Our paper aims to address these two aspects. We propose efficient algorithms for computing $s$-line graphs, apply different heuristics to eliminate redundant work, and discuss the parallelization strategies for our algorithms. To the best of our knowledge, we are the first to take such endeavor and propose efficient parallel algorithms for $s$-line graph computation. Considering the potentials for applicability of $s$-overlap computation in many application domains, we hope our algorithms will be useful to the data analytics and network science community.


\rem{While $s$-overlaps in hypergraphs capture richer set of interactions and commonalities among different entities, with some additional steps we can still leverage existing graph analytics on the new sets of relationships discovered in the $s$-overlap computation step. This entails computing the $s$-overlaps, then creating the new adjacency (where there exists an edge between two hyperedges if they share at least $s$ common vertices), constructing a line graph based on this adjacency (with hyperedges as vertices and edges as said connections between hyperedges) and finally applying optimized graph kernels on the computed line graph. }


The contributions of this paper are as follows:
\begin{itemize}[noitemsep,topsep=0pt]
    \item Identifying the core kernel for hypergraph-related connectivity and traversal algorithms, namely $s$-overlap computation. 
    Efficient $s$-overlap and corresponding  $s$-line graph computation algorithms and parallelization techniques for the $s$-overlap computation algorithms.
    \item Implementation of our own customized cyclic range partitioner to distribute workload cyclically among the threads to avoid load imbalance. We also consider Intel's Threading Building Block's built-in blocked range partitioning strategy and evaluate both cyclic and blocked distribution strategies for our parallel algorithms. 
    \item Experimental results demonstrating the efficacy of our algorithm and heuristics.
\end{itemize}
The rest of the paper is organized as follows. \Cref{sec:motivating} provides a motivating example of an application of $s$-line graph to gain valuable insight about a dataset. In \Cref{sec:background}, we provide the formal mathematical definitions and concepts  related to hypergraphs and our current work. Next, we present our algorithm and heuristics in \Cref{sec:our_algorithm}. We report our experimental results in~\Cref{sec:evaluation}. We  discuss related work in \Cref{sec:related_work} and draw final conclusion in~\Cref{sec:conclusion}. 

\section{{\color{black}A Motivating Example}}\label{sec:motivating}
 \begin{wrapfigure}{l}{.2\textwidth}
 \vspace{-2ex} 
 \includegraphics[width=\linewidth]{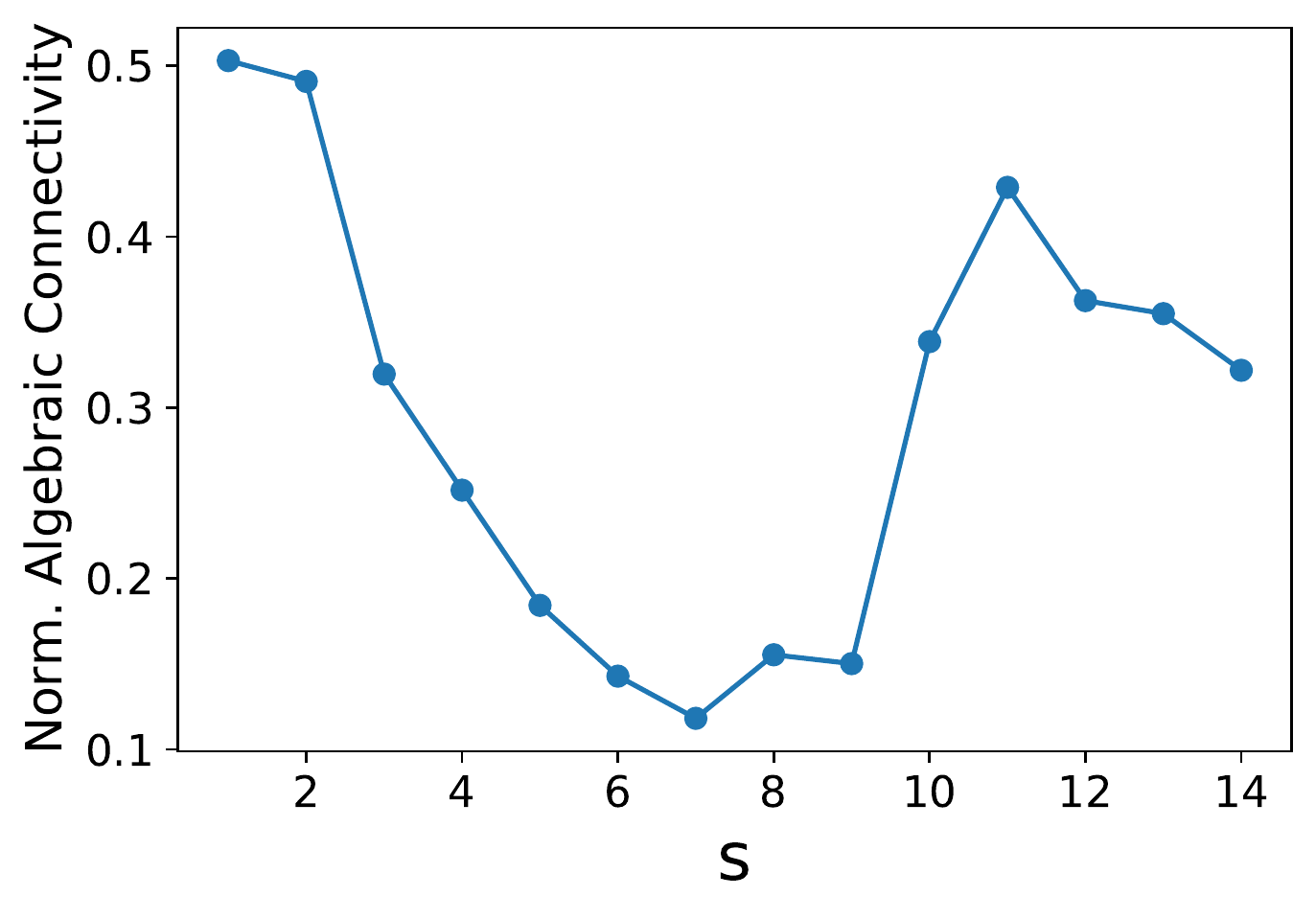}
    \caption{\small \textit {Normalized algebraic connectivity for dataset in ~\cite{DisGeNETref} with various $s$ values.} }\label{fig:algebraic_dis}
 \vspace{-3.2ex}
\end{wrapfigure}    

For certain hypergraph analytics, the formation of $s$-line graphs is strictly necessary. 
To illustrate a particular type of analysis that necessitates $s$-line graph construction, we compute the normalized algebraic connectivity of the $s$-line graphs of a Disease-Gene dataset~\cite{DisGeNETref}. Normalized algebraic connectivity is the second smallest eigenvalue of the normalized Laplacian matrix~\cite{chung1997spectral}; larger values imply stronger connectivity properties of the $s$-line graph and hence the hypergraph. As plotted in \Cref{fig:algebraic_dis}, the fluctuations in algebraic connectivity across values of $s$ reveals information about the network structure: the sharp dip from $s=1$ until $s=7$ suggests most diseases are linked to one another via sparse gene overlaps. The subsequent increase until $s=11$ suggests that diseases which are associated with at least 11 genes are notable in being more well-connected amongst each other. In this way, eigenvalues can provide insight into how well each of the connected components in an $s$-line graph remains connected and consequently provide insight about the original hypergraph connectivity. In addition, as the $s$ value grows, these techniques can assist in understanding how well the connectivity is preserved. These observations are a small example of the insights afforded by spectral methods; see \cite{chung1997spectral} for more.

\section{Background}\label{sec:background}
Since there are some disparities among different definitions of hypergraphs in the literature (in terms of allowing duplicates, isolated vertex etc.), here we adopt the most general definitions from~\cite{aksoy2020hypernetwork}. 

A \textit{hypergraph} $H = (V,E)$ has a set $V = \{v_1,\dots, v_n\}$ of elements called \textit{vertices}, and an indexed family of sets $E = (e_1,\dots, e_m)$ called \textit{hyperedges} in which $e_i \subseteq V$ for $i = 1,\dots, m$. 
The \emph{degree} $d(v)= | e \ni v|$ of a vertex $v$ is the number of hyperedges it belongs to. The \emph{size} (degree) $|e|$ of a hyperedge $e$ is the number of vertices it contains. A hypergraph for which the size of all the hyperedges are the same, let's say $k$, is called a \emph{$k$-uniform} hypergraph. If the size of the hyperedges varies, the hypergraph is \emph{non-uniform}. In case of graphs, $|e_i| = 2$, i.e. an edge connects exactly two vertices. Hence, graphs are also called \emph{2-uniform hypergraph}.

Defining $E$ as an indexed family of sets (as opposed to a set system or a multi-set) allows for multiple copies of edges that can be differentiated by index. This definition also allows for isolated vertices not included in any hyperedge,  empty edges, and singleton edges. In a hypergraph two vertices are considered \textit{adjacent} if there is a hyperedge $e_i$ that contains both of these vertices. We do not allow self-loops in this paper.

For each hypergraph $H = (V,E)$, there exists a {\it dual} hypergraph $H^*$ whose vertices are the edges $e_i$ of $H$ and whose hyperedges are the vertices $v_i$ of $H$. The incidence matrix, $I$ of a hypergraph $H=(V, E)$ is an  $n \times m$ matrix where
{
\setlength\abovedisplayskip{0pt}
  \setlength\belowdisplayskip{0pt}
\begin{equation}
  I(i, j) =
  \begin{cases}
    1 & \text{if $v_i \in e_j$} \\
    0 & \text{otherwise}
  \end{cases}
\end{equation}
}

The \emph{representative} graph, \emph{intersection graph} or \emph{line} graph $L(H)$ of a hypergraph $H$ is a simple graph whose vertices $x_1, \dots, x_m$ represent the hyperedges $e_1, \dots, e_m$ of $H$, and vertex $x_i$ is connected with vertex $x_j$ if $e_i\cap e_j \neq \emptyset$, for $i\neq j$. The edges $\{x_i,x_j\}$ of $L(H)$ can be assigned weights $w_{i,j} = |e_i \cap e_j|$ indicating the size of the hyperedge intersection. For two such nodes $x_i,x_j$, for any $s$ for which $s \leq w_{i,j}$, we say that $e_i$ and $e_j$  have an  \textit{$s-$overlap} or $s-$width. For some $s \ge 1$, restricting $L(H)$ to those edges with weight $w_{i,j} \ge s$ yields the {\it $s$-line graph} denoted $L_s(H)$. Note that thereby the line graph $L(H) = L_1(H)$ is the $1$-line graph. 



Hypergraphs are one-to-one with  \emph{bipartite} graphs by the following construction.  
\begin{wrapfigure}{l}{.2\textwidth}
 \vspace{-2ex}
       \includegraphics[width=\linewidth]{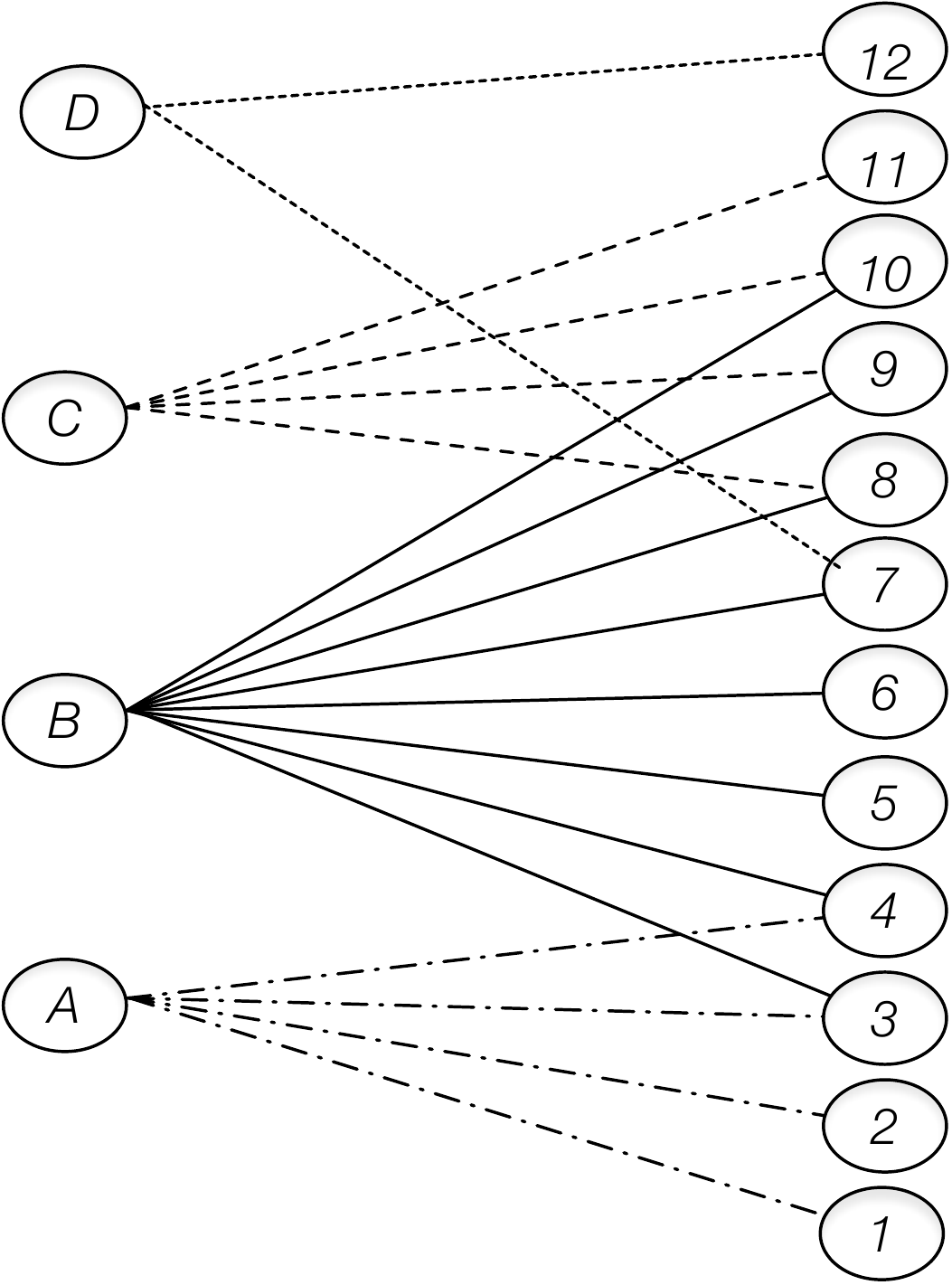}
    \caption{\small \textit { 
    The bipartite graph representation for $H$ in \Cref{fig:hypergraph_example}. } }\label{fig:bipartite_graph}
 \vspace{-3.0ex}
\end{wrapfigure}
A \emph{bipartite} graph consists of two disjoint sets of vertices such that no two vertices within the same set are connected by an edge. The bipartite graph of a hypergraph $H$ has one set of vertices, $V_1$ consisting of the  vertices $V$ and another set $V_2$ consisting of the hyperedges $E$. An edge exists between two vertices of the bipartite graph if the corresponding vertex is included in the incidence of the hyperedge.
\Cref{fig:bipartite_graph} shows 
the bipartite graph for $H$ from \Cref{fig:euler_diagram_hypergraph}.

\newcommand{\tup}[1]{\left< #1 \right>}			


\rem{For a hypergraph $H=(V, E)$, a subset of hyperedges $F \subseteq E$ constitutes an $s$-connected component if there exists at least one $s$-walk between any pair of edges in $F$. }

\rem{and there is no $E_j \subseteq E$ such that $E_i \subset E_j$.}


\section{Algorithms for Computing \texorpdfstring{$s$}{}-overlaps and \texorpdfstring{$s$}{}-line Graphs}\label{sec:our_algorithm}
In this section, we discuss in detail our efficient algorithm and heuristics to compute the $s$-overlaps of a hypergraph. 
Given a dataset represented as a hypergraph $H$, by computing $s$-overlaps, we can infer interesting relationships among different entities in the dataset and subsequently apply any optimized graph algorithm  to the line graph, $L(H)$ by executing the following steps.

\rem{
{
\setlength{\textfloatsep}{0pt}
\small
\begin{algorithm}[t]
\caption{A naive algorithm to compute the $s$-overlaps of a hypergraph and construct the edge list of the line graph based on $s$-overlaps.\\
 \textbf{Input:} Hypergraph $H = (V, E)$, $s$ \\
 \textbf{Output:} $s$-line graph edge list $L_s(H)$}
\label{algo:naive_s_overlap_serial}
\begin{algorithmic}[1]
\State $L_s(H) \gets \emptyset$
  \ForAll{hyperedge $e_i \in E$}\label{naive:outterloop}
    \ForAll{hyperedge $e_j  \in E\setminus \{e_i\} $}\label{naive:innerloop}
    \State count $\gets set\_intersection(e_i, e_j)$
    \If{count $\geq s$}
        \State $L_s(H) \gets L_s(H) \cup  \{e_i, e_j\}$
    \EndIf
    \EndFor
  \EndFor \vspace{-1em}
 \State \textbf{return} $L_s(H)$
\end{algorithmic}
\end{algorithm}
}
}
\rem{To simplify our discussion, we will only consider $s$-overlaps for hyperedges. Similar techniques can be applied when considering $s$-overlaps for vertices too.}
  

\textbf{Step 1:} Compute the number of common neighbors between each pair of hyperedges in $H$ and discard any pair(s) for which the common neighbor count is less than $s$. This is the $s$-overlap computation step.

\textbf{Step 2:} Construct the $s$-line graph $L_s(H)$ based on the $s$-overlap computation. This involves only considering the hyperedge pairs for which there is at least $s$ common neighbors. Each of these edge pairs will constitute an edge in the edgelist of $L_s(H)$. Once we compute all such pairs of hyperedges, we construct the $s$-line graph $L_s(H)$. In $L_s(H)$, each hyperedge $e_i$ is considered as a vertex $vl_i$ and there exists an edge between vertex $vl_i$ and $vl_j$ if there is at least $s$ common neighbors between the corresponding hyperedges $e_i$ and $e_j$ in $H$. For example, \Cref{fig:linegraph} shows the line graphs of the hypergraph in \cref{fig:hypergraph_example} when we consider $s=1,2,3$. For constructing a 2-line graph, here in the original hypergraph $H$ in \cref{fig:hypergraph_example}, the hyperedge pairs $\{B, C\}$ and $\{A, B\}$ shares at least 2 vertices: ($\{8,9,10\}$ for the first pair $\{B, C\}$ and $\{3,4\}$ for the other). Hence, in $L_2(H)$ (\cref{fig:linegraph}),  we include an  edge between $B$ and $C$ and another edge between $A$ and $B$. Note that, there is no edge between $A$ and $C$, since they do not share any vertices ($s<3$ between $A$ and $C$).  

\textbf{Step 3:} Once the $s$-line graph is constructed from the  $s$-overlaps, we can execute any graph algorithm of interest on the $s$-line graph.



\subsection{Naive algorithm to compute \texorpdfstring{$s$}{}-overlaps}
A naive algorithm, which computes the $s$-overlaps and constructs the edge list of the line graph based on the computed $s$-overlaps, 
considers each pair of hyperedges in the hypergraph to see whether there are at least $s$ common vertices shared by the pair. If this is the case, the hyperedge pair $(e_i, e_j)$ is added to the edge list of the line graph. However, many hyperedges may not have any common neighbor (vertex) to connect them. Hence, this algorithm will execute a lot of redundant work when exhaustively checking each pair of hyperedges. 

\renewcommand{\algorithmicforall}{\textbf{for each}}
\newcommand{\visited}{\ensuremath{\mathit{visited}}}
\algrenewcommand\textproc{}
{
\setlength{\textfloatsep}{0pt}
\begin{algorithm}[t]
\small
\caption{An alternative algorithm to compute the $s$-overlaps of a hypergraph and construct the edge list of the line graph based on $s$-overlaps. \\
\textbf{Input:} Hypergraph $H = (V, E)$, $s$ \\
\textbf{Output:} $s$-line graph edge list $L_s(H)$ 
}
\label{algo:efficient_s_overlap_serial}
\begin{algorithmic}[1]
\State $L_s(H) \gets \emptyset$
  \ForAll{hyperedge $e_i \in E$}\label{efficient_serial:outterloop}
     \ForAll{incident vertex $v_k$ of $e_i$}\label{efficient_serial:neighofE}
        \ForAll{incident hyperedge $e_j$ of $v_k$ such that ($j \neq i$)}\label{efficient_serial:neighofN}
            \State count $\gets set\_intersection(e_i, e_j)$
            \If{count $\geq s $}
                \State $L_s(H) \gets L_s(H) \cup  \{e_i, e_j\}$
            \EndIf
        \EndFor
    \EndFor
  \EndFor
  \vspace{-1.5em}
\State \textbf{return} $L_s(H)$
\end{algorithmic}
\end{algorithm}
}

\subsection{An alternative algorithm to compute \texorpdfstring{$s$}{}-overlaps}
To avoid considering redundant pairs of hyperedges when computing $s$-overlaps, we propose an alternative algorithm in~\Cref{algo:efficient_s_overlap_serial}. In this algorithm, instead of considering each pair of hyperedges in the hypergraph $\{(e_i, e_j) | \ e_i \in E, \ e_j \in E, \ i\neq j\}$ for $s$-overlap computation, we only consider those hyperedge pairs $(e_i,e_j)$ that are incident to a common vertex. Mathematically, we only consider hyperedge pairs $\{(e_i, e_j) | \ e_i \in E, \ e_j \in E, \ i\neq j\ \land \exists \ n_k | \ n_k \in V, \ n_k \in I(e_i), \ n_k \in I(e_j)\}$.

{
\setlength{\textfloatsep}{0pt}
\begin{algorithm}[t]
\small
\caption{An efficient algorithm with heuristics to compute the $s$-overlaps of a hypergraph and construct the edge list of the line graph based on $s$-overlaps.\\
\textbf{Input:} Hypergraph $H = (V, E)$,  $size[e_i] \ \forall e_i \in E$, $s$ \\
 \textbf{Output:} $s$-line graph edge list $L_s(H)$
 }
\label{algo:efficient_s_overlap_serial_heuristics}
\begin{algorithmic}[1]
\State $L_s(H) \gets \emptyset$
  \ForAll{hyperedge $e_i \in E$}\label{efficient:outterloop}\label{line:eff_edge_pair}
    \If{$size[e_i$] $< s$} \label{line:deg_prun_ei_start} \Comment{Degree-based pruning}
        \State \textbf{continue} 
    \EndIf \label{line:deg_prun_ei_end}
    \vspace{-0.5em}
    \ForAll{$e_j \in E\setminus$ \{$e_i$\}}
        \State \visited[$e_j$] $\leftarrow$ false
    \EndFor 
    \vspace{-0.5em}
    \ForAll{incident vertex $v_k$ of $e_i$}\label{line:efficient:neighofE}
        \ForAll{incident hyperedge $e_j$ of $v_k$}\label{efficient:neighofN}
        \If{$e_i \geqslant e_j$} \Comment{Consider strictly upper triangular part of the hyperedge adjacency matrix}\label{efficient:skip}
            \State \textbf{continue} 
        \EndIf
        \If{$size[e_j$] $< s$} \label{line:deg_prun_ej_start}
            \State \textbf{continue} 
        \EndIf
        \If{\visited[$e_j$] $==$ true}\label{line:visited_start}
            \State \textbf{continue} \Comment{The current hyperedge $e_j$ has already been considered for the set intersection with $e_i$} 
        \Else
            \State \visited[$e_j$] = true \label{line:visited_end}
        \EndIf
        \vspace{-0.5em}
        \State s\_overlapped $\gets set\_intersection(e_i, e_j, s)$ \label{line:eff_set_intersection}
        \If{s\_overlapped $==$ true}
            \State $L_s(H) \gets L_s(H) \cup  \{e_i, e_j\}$
        \EndIf
        \EndFor
    \EndFor
  \EndFor
  \vspace{-1.5em}
  \State \textbf{return} $L_s(H)$
\end{algorithmic}
\end{algorithm}
}

\subsection{Pruning Strategies for Optimization}
To avoid redundant work, we apply the following pruning techniques in different stages of \Cref{algo:efficient_s_overlap_serial}, which results in \Cref{algo:efficient_s_overlap_serial_heuristics}. The heuristics for pruning redundant work are:
\begin{itemize}[noitemsep,topsep=0pt,leftmargin=*]
    \item \textbf{Degree-based pruning:} When considering a pair of hyperedges $(e_i, e_j)$ for $s$-overlap computation, since it is required that there are at least $s$ vertices in the incidence list of each of these hyperedges,  if the size of either edge  $|e_i|$ or $|e_j|$ is smaller than $s$, we can exclude the hyperedge from consideration when computing the  $s$-overlap~(\Lineref{line:deg_prun_ei_start} and \Lineref{line:deg_prun_ej_start} in \Cref{algo:efficient_s_overlap_serial_heuristics}). 
    
    \item \textbf{Skip already visited hyperedges:} Consider a connected hyperedge pair $(e_i, e_j)$. Here $e_i$ and $e_j$ denote hyperedge IDs (associated with each set of vertices in $e_i$ and $e_j$). It is possible that, $e_i$'s one-hop hyperedge neighbor, in this case $e_j$, can be reached via multiple different vertices $v_k, v_l, v_p$ etc., where $k\neq l\neq p$ (\Lineref{line:efficient:neighofE} in \Cref{algo:efficient_s_overlap_serial_heuristics}). This means there exist ``wedges'' $(e_i, v_k, e_j)$, $(e_i, v_l, e_j)$, $(e_i, v_p, e_j)$ etc. in the bipartite graph representation of the hypergraph. Since we only need to perform set intersection between potential hyperedge pairs once, we can keep track of whether a one-hop hyperedge $e_j$ has already been considered for intersection by maintaining a \texttt{visited} array. This boolean \texttt{visited} array of size $m$ is allocated outside of the outer loop in \Cref{algo:efficient_s_overlap_serial_heuristics}. At the beginning of each iteration of the outer loop, the array is cleared first. During the execution of the innermost loop, we check whether the current hyperedge $e_j$  has already been considered for set intersection. If this hyperedge is already visited, we can skip the set intersection. Otherwise we set the $visited$ entry to true for $e_j$ and proceed with the set intersection~(\linesref{line:visited_start}{line:visited_end} in \Cref{algo:efficient_s_overlap_serial_heuristics}).
    
    \item \textbf{Short-circuiting in set intersection:} During the set intersection of the vertex neighbor lists of hyperedges $e_i$ and $e_j$, as soon as we detect $s$  overlaps of vertices between $e_i$ and $e_j$, we can return immediately, without iterating through the complete list of neighbors, and return a boolean value of \texttt{true}, instead of intersection \texttt{count}, indicating that the edge pair has at least $s$ common neighbors. Otherwise the \texttt{set\_intersection} returns false. The modified set intersection takes $s$ as an input for comparison. 
    
    
    \item \textbf{For set intersection, consider strictly upper triangular part of the hyperedge adjacency:} Consider a \emph{connected} hyperedge pair $(e_i, e_j)$. Since each edge $e_i$ and $e_j$ will be considered independently (on ~\Lineref{line:eff_edge_pair} of \Cref{algo:efficient_s_overlap_serial_heuristics}) , and since the connected  pair has at least one neighboring vertex in common, set intersection will be performed twice with the same edges $(e_i, e_j)$ individually on ~\Lineref{line:eff_set_intersection}: by considering the  wedges $(e_i, n_k, e_j)$ and $(e_j, n_k, e_i)$, where $n_k$ is a common neighboring vertex. We can eliminate one of these redundant set  intersections by only considering wedge $(e_i, n_k, e_j)$ when $e_i > e_j$. In other words, we only consider the upper triangular matrix of edge adjacency (here the edge adjacency refers to the $m \times m$ matrix where a non-zero entry exists whenever the edge-pair ($(e_i, e_j)$) is at least 1-connected. Note that we are not constructing the edge adjacency concretely, it is just for conceptual reference. Additionally, to avoid self-loops in $L_s(H)$ (when $e_i=e_j$), we only consider strictly upper triangular part of the hyperedge adjacency.
    
\end{itemize}

\subsection{Linear Algebraic Formulation of the  \texorpdfstring{$s$}{}-line graph Computation}
Conceptually, considering the incidence matrix $H$, and identity matrix $I$, $s$-line graph computation can be expressed as the multiplication of two sparse matrices  $H$ and $H^T$ (where $H^T$ is the transpose of matrix $H$), then subtracting $2I$ from it (to avoid self loops in the resultant matrix), and finally  applying a \emph{filtration} operation to each entry of the  $HH^T-2I$ matrix, such that, in the resultant matrix $R$, $R_{i, j} = 0$ if $\{HH^T-2I\}_{i,j} < s$, otherwise $R_{i, j} = 1$.

{\color{black}Computing the sparse general matrix-matrix multiplications (SpGEMM) ~\cite{gustavson_1978_two} and then applying the filter operation to find the edgelist of a $s$-line graph is closely-related (NOT equivalent) to the alternative approach~\Cref{algo:efficient_s_overlap_serial}. However, SpGEMM is time-consuming (results shown later) and there are three reasons why it is not efficient for $s$-line graph computation. First, it considers both the upper triangular and lower triangular parts of the hyperedge adjacency matrix, even though the matrix is symmetric. Second, since SpGEMM is more general, SpGEMM has to compute and materialize the product matrix before applying filtration upon it. This requires extra space to store the intermediate results (while the alternative approach does not). This is the only difference between the alternative approach and SpGEMM plus filtration. Third, it cannot apply heuristics to speedup the computation, such as pruning or short circuiting the set intersection.}

\subsection{Hypersparsity and ID Squeezing}
The edge list of the line graph based on the $s$-overlap computation, $L_s(H)$, contains the original hyperedge IDs. However, as we increase the value of $s$, where $s$ is bounded by the maximum size of the hyperedges, 
many hyperedges may not be included in the edge list of the line graph, simply because there may not be enough overlaps between any such pairs of hyperedges. 
In terms of adjacency/incidence matrix representation, \emph{hypersparsity} refers to the cases where many rows of the matrix do not have any non-zero element (or zero-columns for the transposed matrix). 

Allocating memory and constructing the line graph based on the original IDs of the hyperedges is infeasible due to this observed hypersparsity phenomena. To avoid this problem, we construct a new set of contiguous IDs for the hyperedges (i.e. for the vertices in the line graph) included in the line graph's edge list and maintain a mapping between original IDs and the new IDs. Remapping the IDs is termed as \emph{squeezing}. ID squeezing is an essential step before materializing the line graph adjacency from the edge list. 

\subsection{Parallelization}
We implemented our algorithms in \texttt{C++17} and utilize Intel's Threading Building Block (TBB)~\cite{tbbrepo} to parallelize our algorithms. For example, the outer \texttt{for} loop (\Lineref{efficient:outterloop} in \cref{algo:efficient_s_overlap_serial_heuristics}) is parallelized with TBB's \texttt{parallel\_for}. We invoke TBB's \texttt{parallel\_for} in the form of \texttt{(range, body, partitioner)}. Considering this form enables us to provide custom range to TBB. We discuss one of the custom ranges (cyclic range) in the context of workload balancing techniques later in this section. \texttt{parallel\_for} breaks the hyperedge iteration space (range) in chunks and schedules each chunk on a separate thread. Each thread executes the body (a lambda function, containing the logic for s-line computation) on each of these chunks.

\subsubsection{Workload distribution strategies for load balancing.} The first argument (range) of TBB's \texttt{parallel\_for} API, \texttt{(range, body, partitioner)}, provides provision to supply different strategies to partition and distribute the iteration space (range) among the threads. As long as the provided range adheres to the C++ \texttt{Range} concept~\cite{niebler2018one}, it enables the user to experiment with different workload balancing techniques.

\textbf{Blocked range.} When considering one-dimensional iteration space of hyperedges for partitioning, one possibility is to specify the range of hyperedges as a \texttt{blocked\_range} (other options include 2D, 3D partitioning etc.). A \texttt{blocked\_range} represents a recursively splittable range and is provided as a built-in range in TBB. Each thread is assigned a contiguous chunk of iteration space to work on. Additionally we  specify TBB's \texttt{auto\_partitioner} as the range partitioning strategy to enable optimization of the parallel loop by range subdivision based on work-stealing events. The \texttt{auto\_partitioner} adjusts the number of chunks depending on the available execution resources and load balancing needs.

However, applying blocked range strategy to distribute the workload can be problematic for applications with irregular work pattern such as hypergraph algorithms. This is because the degree distributions of the hyperedges is mostly non-uniform~(\cref{fig:degree_distribution}). Splitting the hyperedge range (ID range of the hyperedges) with blocked distribution can create uneven workload if hyperedges with consecutive IDs have high degrees. In this case, some threads will be assigned larger workload compared to others.

\textbf{Cyclic range.} An alternative to blocked\_range strategy is cyclic range, where, assuming the stride size is equal to the number of threads $nt$, thread 0 processes hyperedges $e_0, e_{0+nt}, e_{0+2*nt}, e_{0+3*nt}$ etc., thread 1 processes hyperedges $e_1, e_{1+nt}, e_{1+2*nt}, e_{1+3*nt}$ and so on. Here $e_i$ denotes hyperedge ID. If a hypergraph contains high-degree hyperedges with consecutive IDs, cyclic range will distribute the workload more evenly among the threads than the blocked range by reducing the effect of consecutive high-degree hyperedge IDs (induced locality) on the workload.


\textbf{Relabel by degree.} When applying cyclic workload distribution strategy, it may be useful to relabel the hyperedge IDs according to their degrees, so that high-degree hyperedges will be given smaller IDs and low-degree hyperedges will obtain larger IDs. The high-degree hyperedges will be adjacent to each other (since their new IDs will be consecutive). In this way, when processing the hyperedges, each thread will have a better chance of load-balanced workload assignment when applying cyclic distribution.

\begin{figure}[tp]
  \centering
  \includegraphics[width=\linewidth]{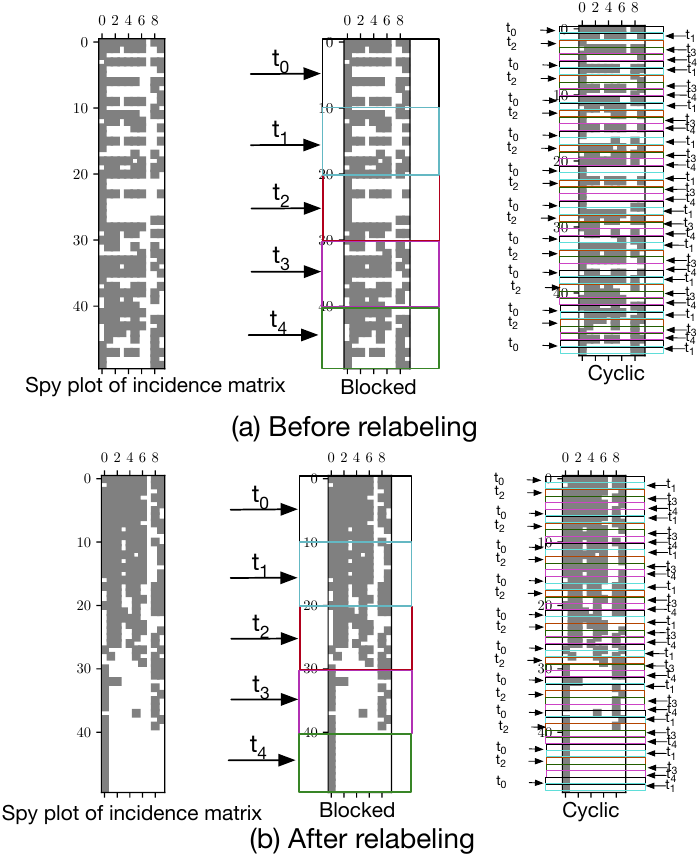}
  \caption{\small \textit {Spy plot of the incidence matrix of a hypergraph before and after relabeling. We also show how workloads are distributed among threads when using blocked vs cyclic range in each case (assuming that workloads are distributed among 5 threads).} }
  \label{fig:relabeling_spyplot}
\end{figure}

~\Cref{fig:relabeling_spyplot} shows the spyplot of a hypergraph incidence matrix before and after relabeling the IDs of the hyperedges. In addition, the figure also illustrates how workload is distributed among threads when considering blocked and cyclic partitioning. As can be observed from the figure, before relabeling the IDs of the hyperedges, blocked range-based partitioning suffers from uneven workload distribution among the threads (color coding in the figure shows the range of hyperedges assigned to each thread for processing). On the other hand, cyclic range circumvents this problem by trying to scatter high-degree hyperedges among all the threads. To make this approach even more effective, relabel-by-degree ensures that all high-degree hyperedges ends up with consecutive IDs so that each thread will choose one high-degree hyperedge ID at a time for processing when considering cyclic distribution of the hyperedges among the threads.

\textbf{Thread-local data structures.} To avoid any duplicate set intersection, each thread also maintains a \texttt{boolean} \texttt{visited} array of size $m=|E|$ to track the set of hyperedges that have already participated in the set intersection with a hyperedge. In addition, each thread also maintains a thread-local \texttt{vector} for inserting edge pairs, if such pair meets the $s$-overlap requirement. Once all the threads finish processing their assigned edge range, the thread-local edgelists are consolidated into one array to construct $L_s(H)$.  

\subsection{Time Complexity Analysis}
Given a hypergraph $H(V,E)$, where $|V|=n$, $|E|=m$, let us denote the average degree of $v_i \in V$ by $\overline{d}$, and average size of $e_i \in E $ by $\overline{k}$.
In  the worst case scenario, each set intersection will perform $O(m)$ comparisons. 
Hence, \Cref{algo:efficient_s_overlap_serial} takes $O(nm^3)$ time in the worst case.

In the worst case, without any heuristic, our efficient algorithm (\Cref{algo:efficient_s_overlap_serial_heuristics}) takes $O(nm^3)$ time. To factor in the effect of the heuristics on the execution time, let us assume that, on average, each set intersection operation performs $\overline{k}$ number of comparisons. Considering this, in the average case, our algorithm takes $O(m*\overline{k}*\overline{d}*\overline{k})$ time. 
The amount of work that degree-based pruning saves ($O(\overline{d}*\overline{k})$) is based on the value of $s$. Since this heuristic is encoded in both the outer loop and inner loop, it saves most of the redundant work by avoiding execution of two innermost loops.
With the upper triangular heuristic, it saves half of the work by skipping one endpoint of every hyperedge pair. Additionally, with short-circuiting in set intersection, we only need $s$ successful  comparisons for early termination. 
Our approach takes $O(m*\overline{k}*\overline{d}*\overline{k})$ time after all heuristics kick in.

\section{Evaluation}\label{sec:evaluation}
In this section, we report and analyze our experimental results for computing $s$ overlaps and line graphs.

\subsection{Experimental setup}
For experimental evaluation, we ran our experiments on a machine with a two-socket Intel Xeon Gold 6230 processor, with 20 physical cores per socket, each running at 2.1 GHz, has 28 MB L3 cache. The system has 188 GB of main memory. We implemented our code in C++17, compiled the library with Intel TBB 2020.2.217, GCC 10.1.0 compiler and \texttt{-Ofast} compilation flag. 

 {
\setlength{\tabcolsep}{12pt}

\begin{table}
\footnotesize
  \centering
  \small
  \tabcolsep=0.05cm
  \begin{tabular}{*{7}{c}} \toprule
    Network type & hypergraph & $|V|$ & $|E|$ & $\overline{d}_e$ & $\Delta_{v}$ & $\Delta_{e}$ \\ \midrule
     \multirow{2}{*}{Social} & com-Orkut  & 2.32M & 15M & 7 & 2958 & 9120\\
                                     & Friendster  & 7.94M &1.62M & 14 & 1700 & 9299 \\ 
                                     & Orkut-group & 2.78M & 8.73M & 37 & 40k & 318k \\
                                     & LiveJournal & 3.2M & 7.49M & 15 & 300 & 1.05M \\
                                     \hline
    \multirow{1}{*}{Webgraph} & Web & 27.7M & 12.8M & 11 & 1.1M & 11.6M \\ \hline
    \multirow{1}{*}{Cyber} & activeDNS (256 files) & 4.5M & 43M & ~8 & 1M & 10M\\ \hline
    \multirow{1}{*}{Collaboration} & IMDB & 896k & 3.8M & 8 & 1.6k & 1334\\ \hline
     \multirow{1}{*}{Miscellaneous} & email-EuAll & 265.2k & 265.2k & 1.6 & 7.6k & 930 \\
    \bottomrule
    \end{tabular}
    \caption{\small \textit {Input hypergraph characteristics. Total number of vertices ($|V|$) and hyperedges ($|E|$) along with the average size of each edge ($\overline{d}_{e}$), max degree of a vertex ($\Delta_{v}$), and maximum size of a hyperedge ($\Delta_{e}$) for the hypergraph inputs are tabulated here.} }
    \label{tab:input_hypergraph_prop}
\end{table}
  }
\begin{figure}[tp]
  \centering
  \begin{subfigure}[t]{0.2\textwidth}
    \centering
  \includegraphics[width=\linewidth]{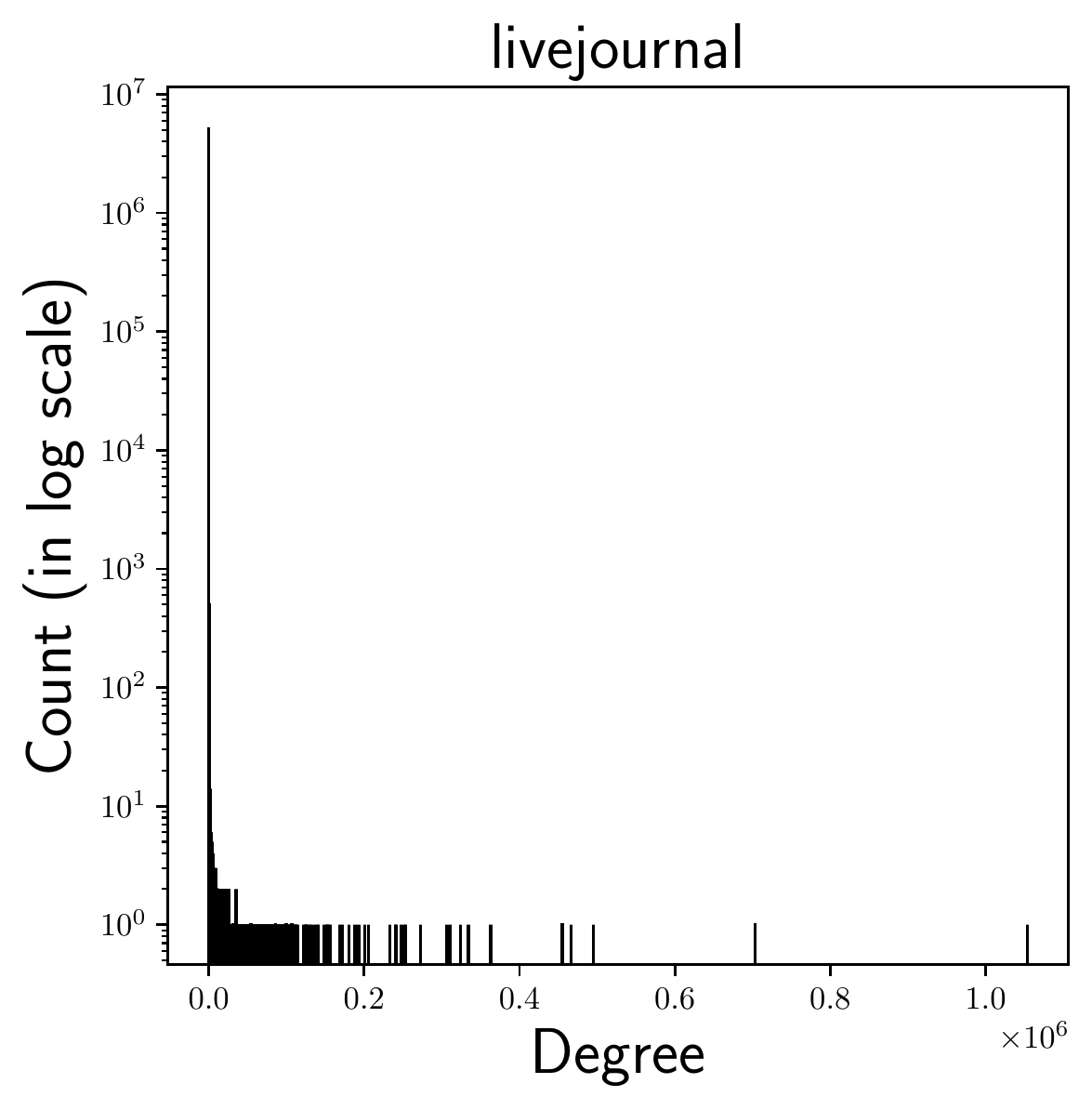}
  \label{fig:degree_livejournal}
  \end{subfigure}
\begin{subfigure}[t]{0.2\textwidth}
    \centering
  \includegraphics[width=\linewidth]{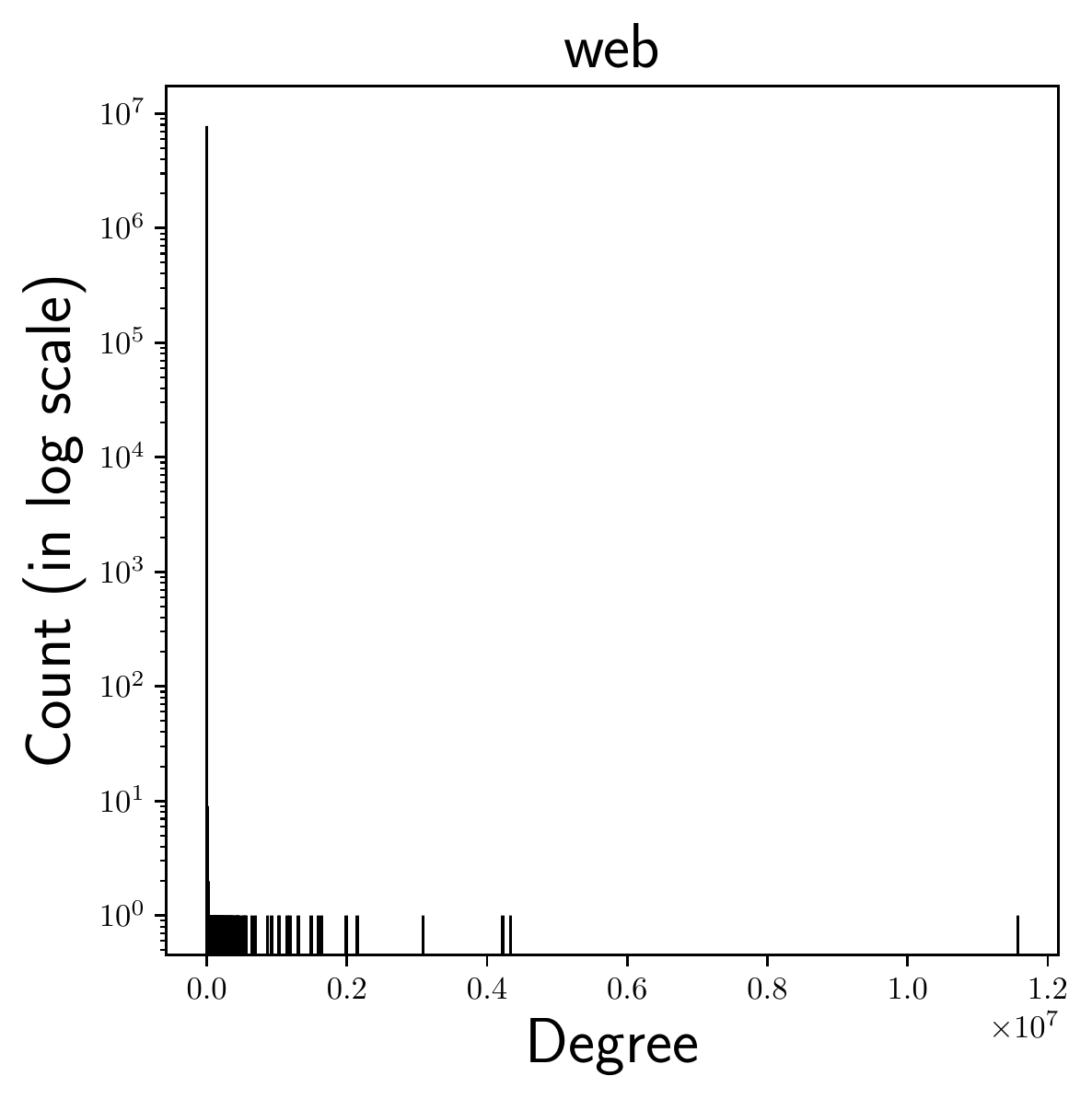}
  \label{fig:web}
  \end{subfigure} \\
  \begin{subfigure}[t]{0.2\textwidth}
    \centering
  \includegraphics[width=\linewidth]{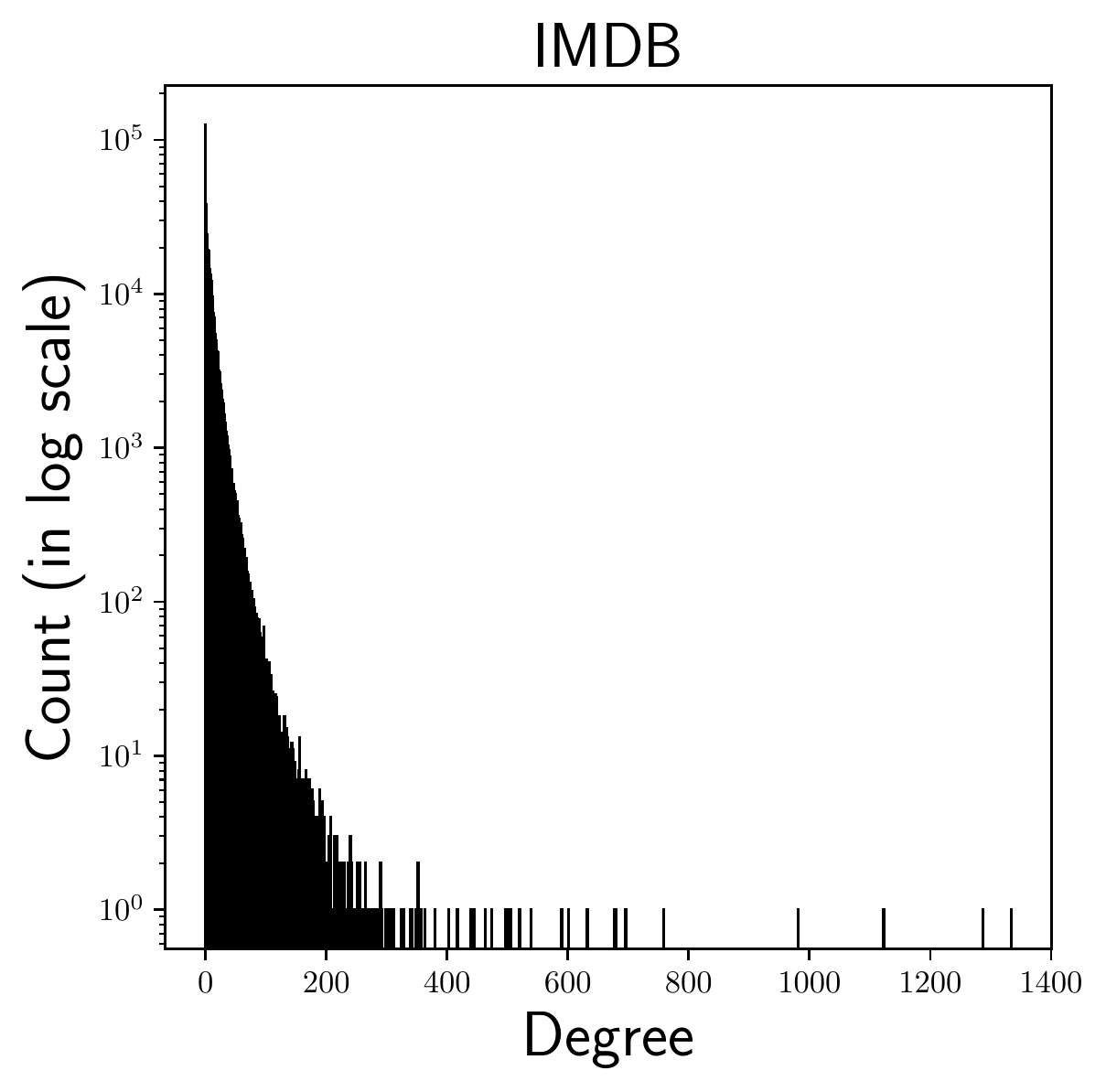}
  \label{fig:degree_IMDB}
  \end{subfigure}
\begin{subfigure}[t]{0.2\textwidth}
    \centering
  \includegraphics[width=\linewidth]{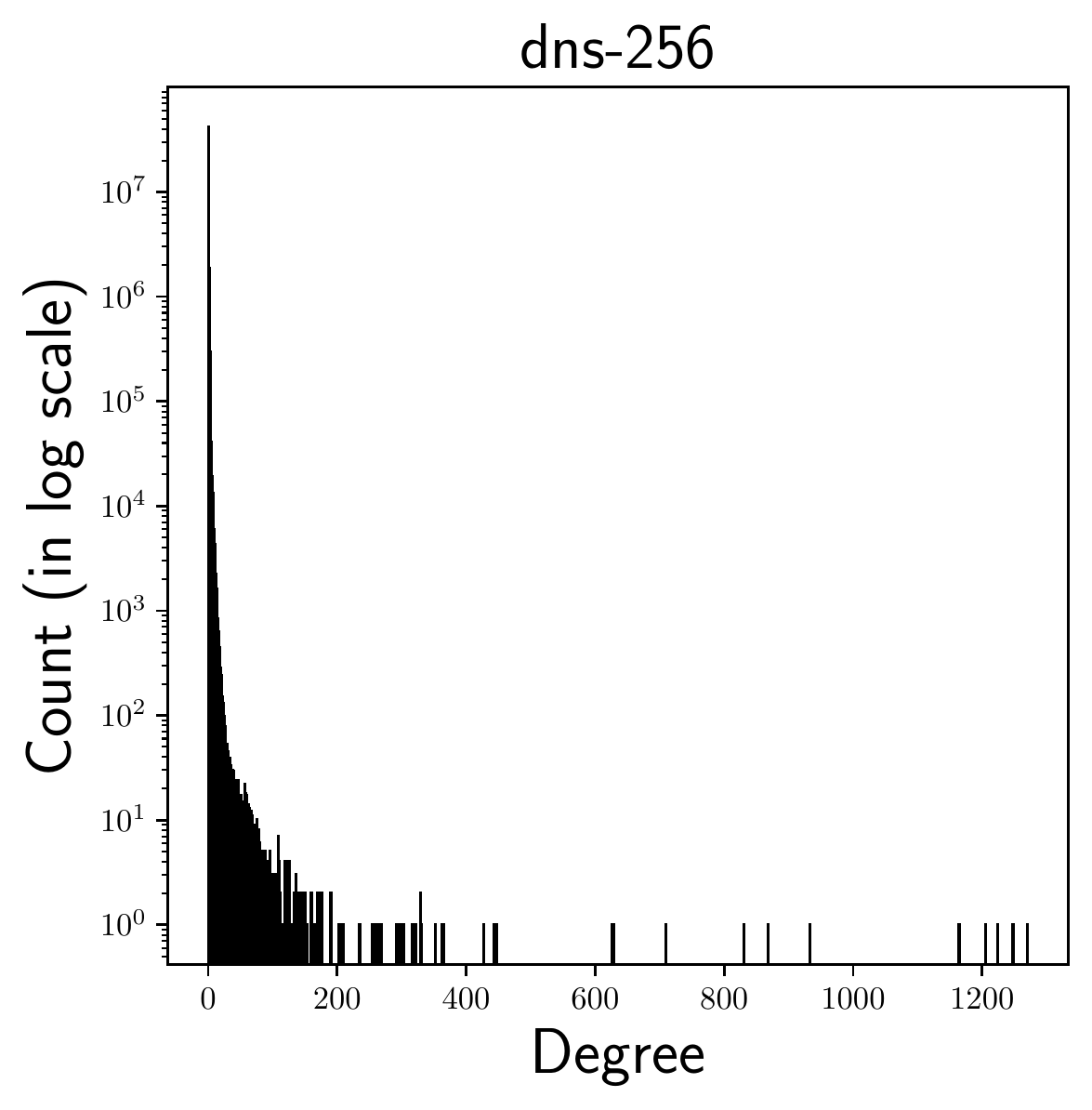}
  \label{fig:dns-256_deg}
  \end{subfigure}
  \vspace{-1em}
 \caption{\small \textit{Hyperedge degree distributions of a representative set of hypergraphs.} }\label{fig:degree_distribution} 
 \vspace{-1.6em}
\end{figure}

\subsection{Datasets}
We conducted our experiments with datasets mentioned in
(\Cref{tab:input_hypergraph_prop}). 
These datasets represent a set of hypergraphs constructed from diverse domains: ranging from social networks to cyber data to web, each having different structural properties. Basic statistics of these hypergraphs are presented in~\Cref{tab:input_hypergraph_prop} and edge degree distributions are shown in~\Cref{fig:degree_distribution}.

The activeDNS (ADNS) dataset from Georgia Institute of Technology~\cite{ActiveDNSdataset} contains mappings from domains to IP addresses. When constructing hypergraphs with ADNS dataset, we consider the domains as the hyperedges and IPs as vertices. We followed the same procedure as described in~\cite{joslyn2020hypergraph} to curate this dataset. 

We also ran our experiments with datasets curated in~\cite{shun_practical_2020}. For these curated datasets, in particular, each hypergraph, constructed from the social network datasets such as com-Orkut and Friendster~\cite{leskovec2016snap} in \Cref{tab:input_hypergraph_prop}, are materialized by running a community detection algorithm on the original dataset from Stanford Large Network Dataset Collection (SNAP)~\cite{leskovec2016snap}. In the resultant hypergraphs, each community is considered as a hyperedge and each member of a community as a vertex. Other larger datasets include orkut-groups, Web, and LiveJournal, collected from Koblenz Network Collection (KONECT)~\cite{kunegis_konect_2013} as bi-partite graphs. The IMDB dataset~\cite{suitsparsecollection} represents a bi-partite graph, where the hyperedges are movies and vertices are actors/actresses. 

\begin{table}[h]
\centering
\begin{tabular}{*{2}{c}}
\toprule
Notation       & Algorithm \\
\midrule
$f0$ & All heuristics included  (\Cref{algo:efficient_s_overlap_serial_heuristics}) \\
$f1$ & Only degree-based pruning \\
$f2$ & Only skipping of the visited hyperedges heuristic \\
$f3$ & Only short-circuiting in the set intersection \\
$f4$ & Baseline efficient with no heuristic (\Cref{algo:efficient_s_overlap_serial}) \\
\bottomrule
\end{tabular}
\caption{\small \textit{ Notation for different algorithms and heuristics.} }\label{table:algo_notation}
\vspace{-1.6em}
\end{table}

\subsection{Experimental Results}
In this section, we discuss the experimental results in detail. First we report the performance improvement of our algorithm over the naive one, as we add different heuristics, one at a time, as well as combining them altogether. Next, we assess the effect of our proposed heuristics as we vary the size of $s$. We also report the scalability results (strong and weak scaling) of our algorithms with different workload distribution strategies. 
We discuss the inputs for which relabel-by-degree with cyclic workload distribution is helpful. Finally we show how the workload is distributed among threads with different workload distribution techniques.   
For the subsequent discussion, we refer to \Cref{table:algo_notation} for the shorthand notations of our different algorithms and heuristics. To be consistent and fair, we applied triangular heuristic across all the algorithms, so that only upper triangular part of the hyperedge adjacency is considered when computing the $s$-overlaps (thus avoiding inclusion of self-loops and duplicate edges in the line graph edge list).


\begin{figure}[tp]
  \centering
  \includegraphics[width=\linewidth]{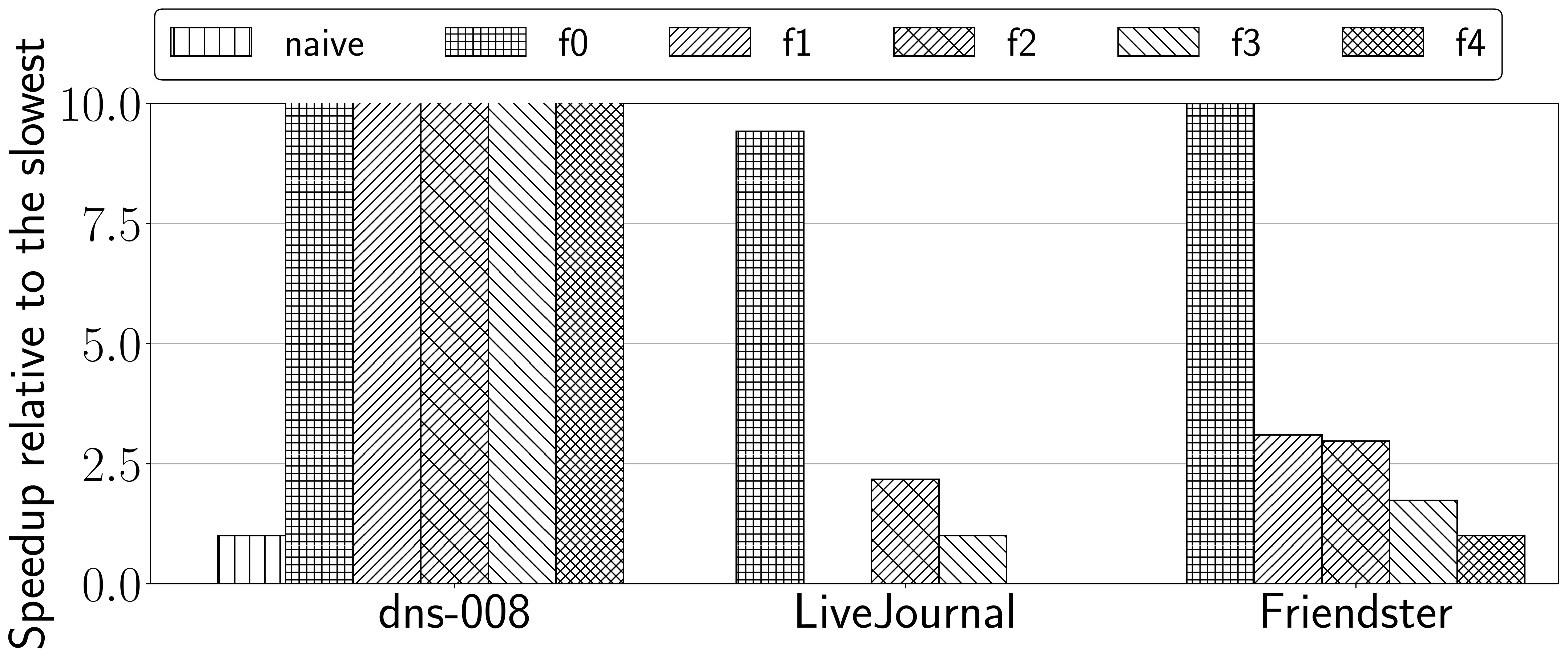}
  \caption{\small \textit{ Performance comparison of different $s$-overlap computation algorithms, including the naive algorithm. Here, for each dataset, we report the speedup relative to the slowest execution time that have ran to completion. In two cases (LiveJournal and Friendster), naive algorithm did not finish execution in a reasonable time. In addition, for LiveJournal, only degree-based pruning heuristic algorithm  did not finish in a reasonable time. Hence these three data points are missing in the plot. The results are reported with 32 threads and for $s=8$.} } 
  \label{fig:speedup_hypergraphs}
\vspace{-1.5em}
\end{figure}

\begin{figure*}[tp]
  \centering
  \begin{subfigure}[t]{0.32\textwidth}
    \centering
  \includegraphics[width=\linewidth]{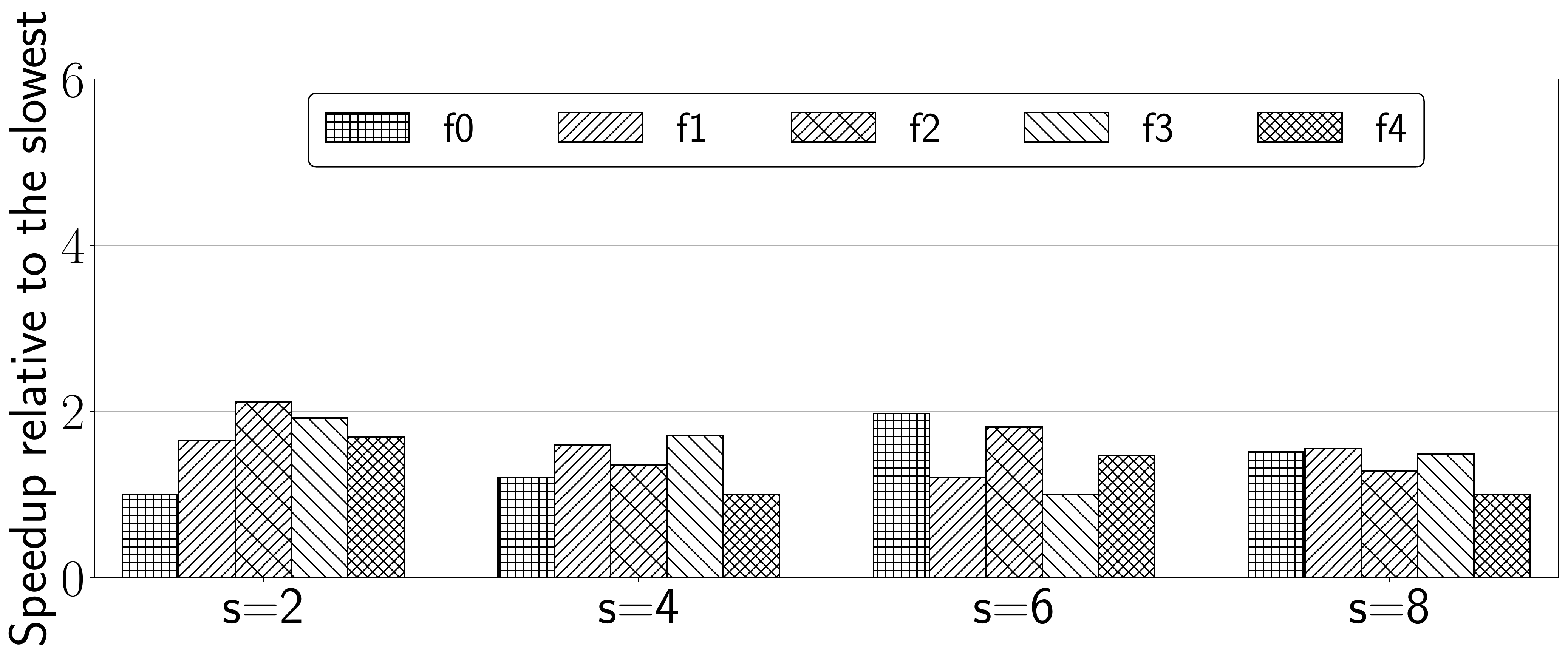}
  \caption{\small \textit{ IMDB. 
  } }
  \label{fig:speedup_imdb}
  \end{subfigure} 
    \begin{subfigure}[t]{0.32\textwidth}
    \centering
  \includegraphics[width=\linewidth]{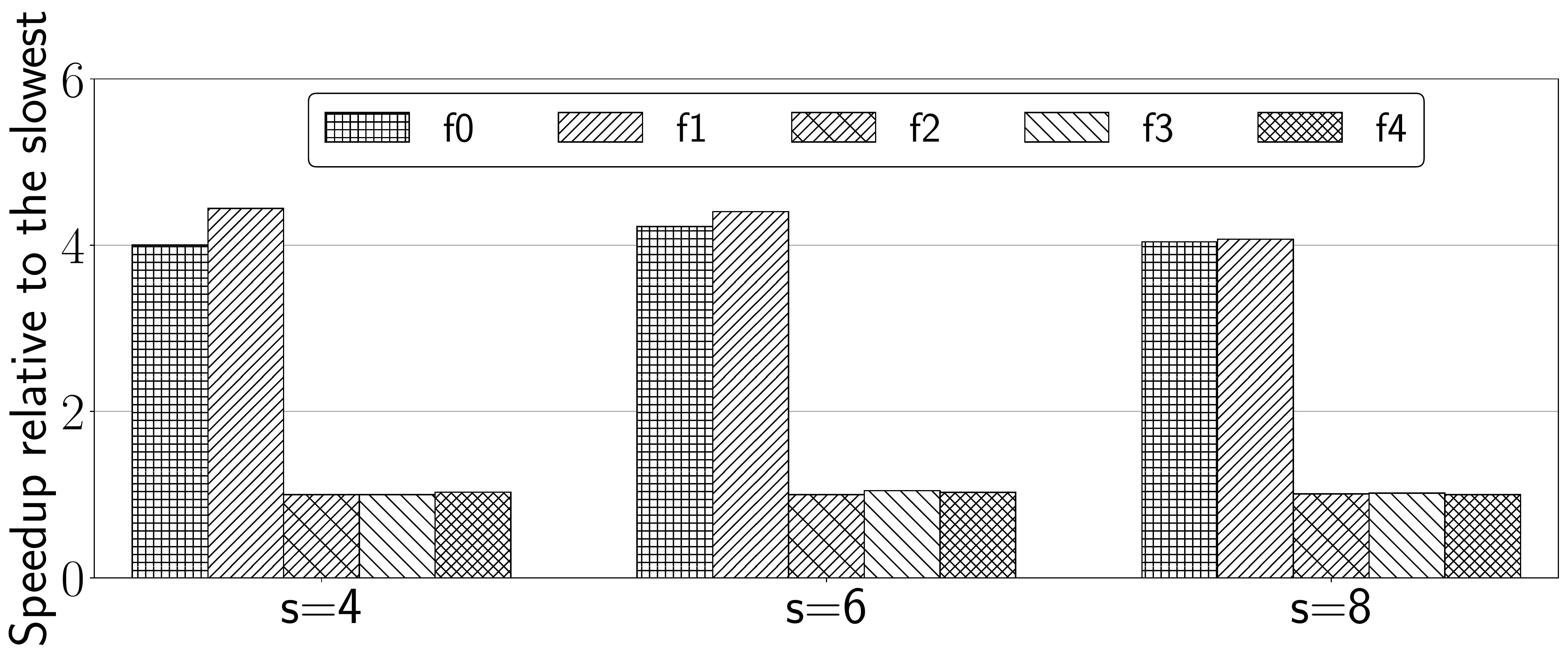}
  \caption{\small \textit{ dns-128. Degree-based pruning is the most impactful  heuristic in this case.} }
  \label{fig:speedup_dns128}
  \end{subfigure} 
  \begin{subfigure}[t]{0.32\textwidth}
    \centering
  \includegraphics[width=\linewidth]{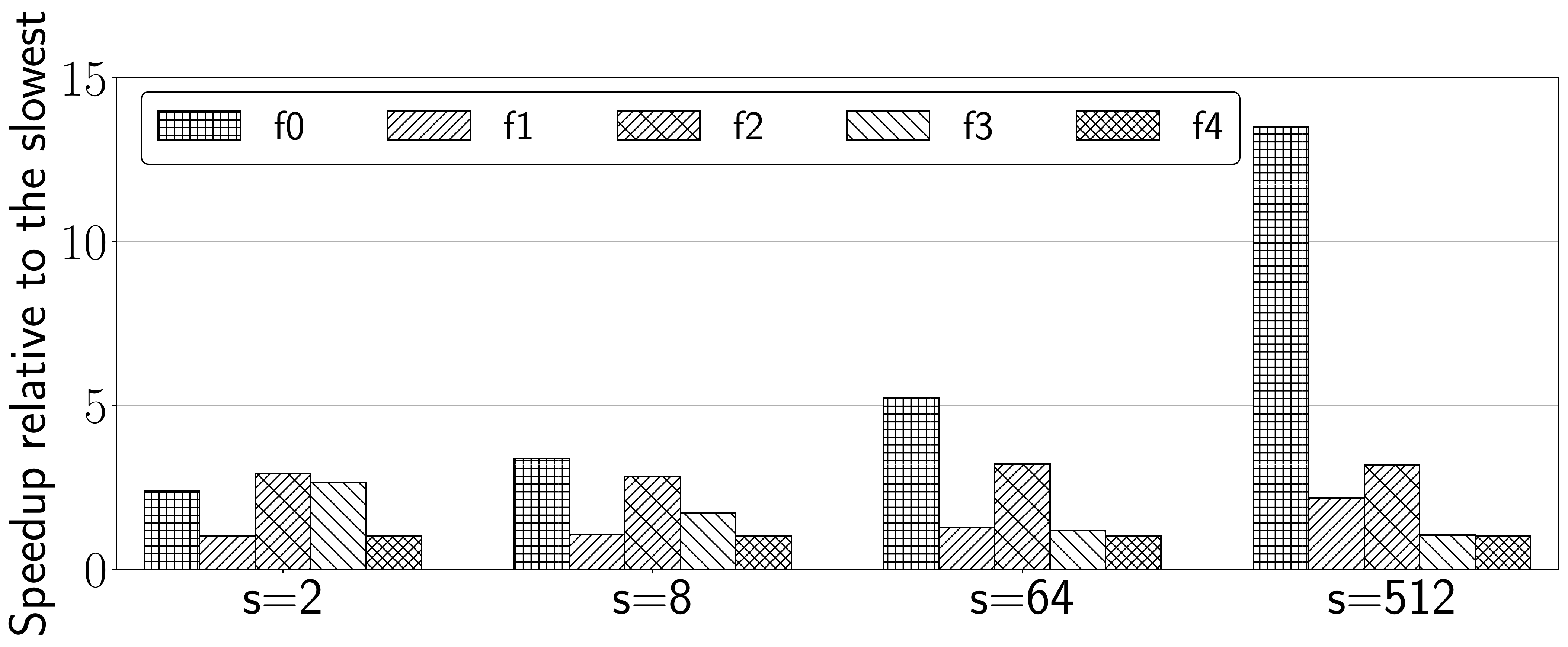}
  \caption{\small \textit{ Friendster. Skipping visited hyperedges heuristic is the most impactful heuristic as well as degree-based pruning for $s>=8$. 
  } }
  \label{fig:speedup_friendster}
  \end{subfigure}\\
 \caption{\small \textit{ Effect of different heuristics on different datasets and with different $s$ values. The speedup is reported by normalizing w.r.t the slowest runtime of all different variations of the algorithm. Here we report execution time running with 32 threads.} }\label{fig:effect_of_heuristics} 
  \vspace{-1.2em}
\end{figure*}

\begin{figure}[ht]
  \centering
  \begin{subfigure}[b]{0.24\textwidth}
    \centering
  \includegraphics[width=\linewidth]{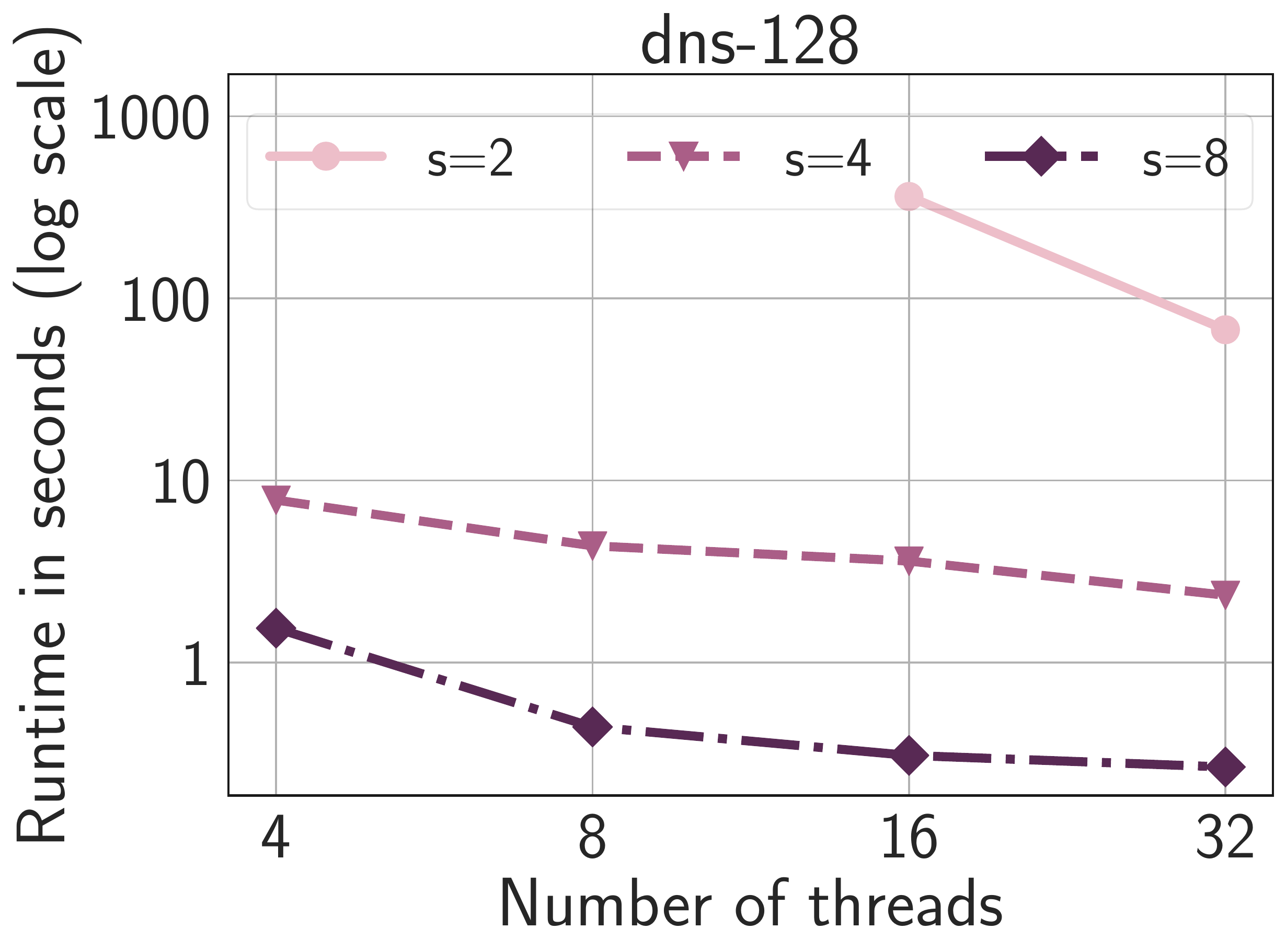}
  \caption{dns-128.}
  \label{fig:dns128_cyclic}
  \end{subfigure} 
 \begin{subfigure}[b]{0.24\textwidth}
    \centering
  \includegraphics[width=\linewidth]{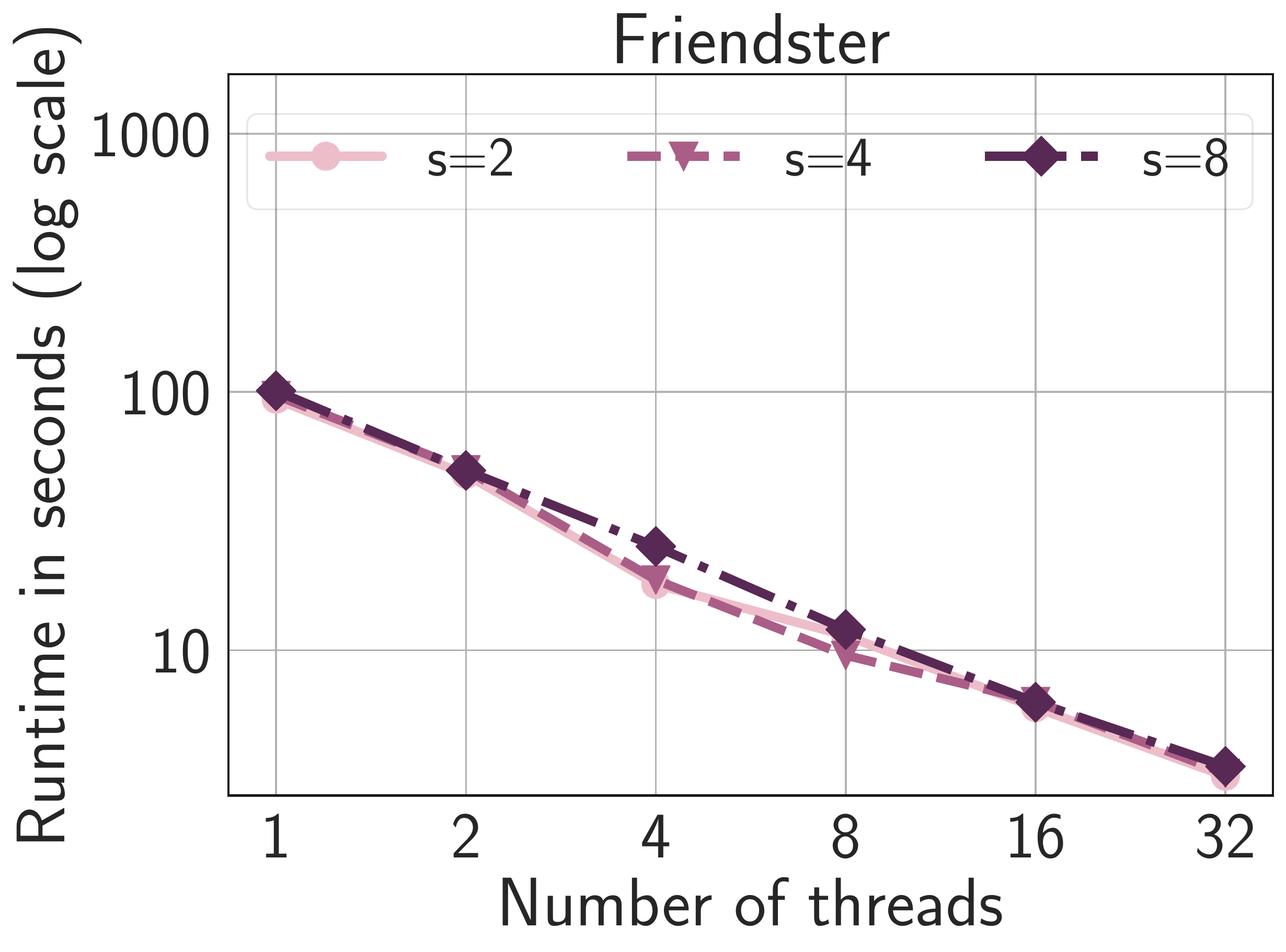}
  \caption{Friendster.}
  \label{fig:friendster_cyclic}
  \end{subfigure} \\
  \begin{subfigure}[b]{0.24\textwidth}
    \centering
  \includegraphics[width=\linewidth]{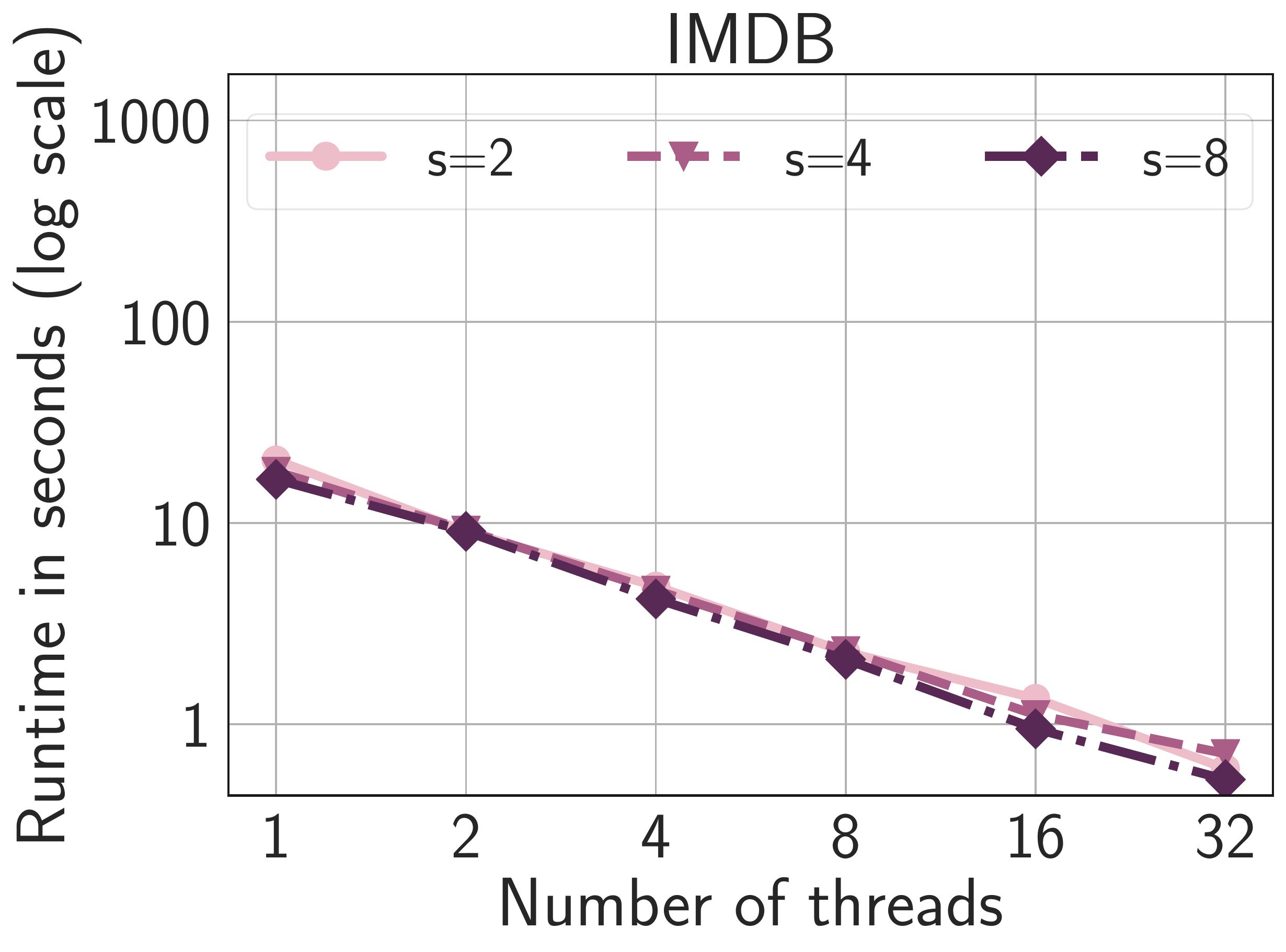}
  \caption{IMDB.}
  \label{fig:imdb_cyclic}
    \end{subfigure} 
    \begin{subfigure}[b]{0.24\textwidth}
    \centering
  \includegraphics[width=\linewidth]{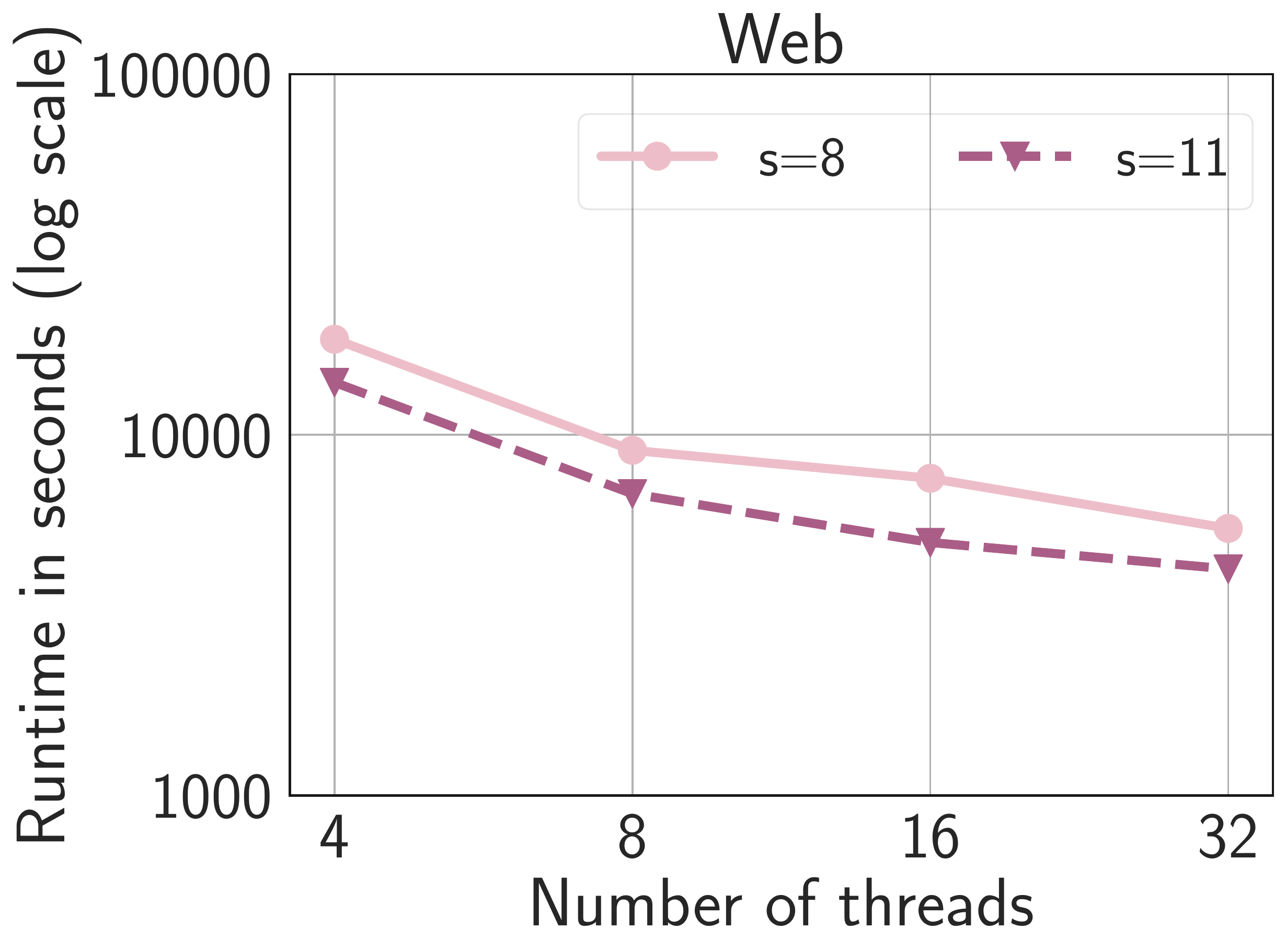}
  \caption{Web.}
  \label{fig:web_cyclic}
  \end{subfigure} 
\vspace{-0.1em}
  \caption{\small \textit{  Strong scaling results with cyclic distribution for \Cref{algo:efficient_s_overlap_serial_heuristics}, $f0$} }
  \label{fig:strong_scaling_cyclic}
  \vspace*{-1.5em}
\end{figure}


\subsubsection{Performance comparison of our algorithm with different heuristics and the naive algorithm}
\Cref{fig:speedup_hypergraphs} reports speedup of our algorithm with different heuristics compared to the naive algorithm, discussed in \Cref{sec:our_algorithm}. Here we only consider  $s=8$. With other $s$ values, the algorithms exhibit similar trend. In these experiments, we consider dns-008, LiveJournal and Friendster datasets. We have chosen these datasets to represent different application domains and varying hypergraph sizes (ranging from medium to large).  


From the figure, we observe that, with dns-008 dataset, our efficient algorithm is significantly faster compared to the naive algorithm. With LiveJournal and Friendster datasets, the naive algorithm could not finish within a reasonable time limit. With dns-008 dataset, our algorithm is more than 10x faster compared to the naive algorithm. Although not visible in~\Cref{fig:speedup_hypergraphs}, we also observed that, for these datasets, the most important heuristic for $s$-overlap computation is the degree-based pruning heuristic ($f1$). In fact, with dns-008 dataset, by only using degree-based pruning heuristic, our algorithm is slightly faster than $f0$. This is because dns-008 dataset has less than 1\% hyperedges with degrees greater than 8. In this case, adding other heuristics introduced more overhead due to additional checks. 

With LiveJournal and Friendster social network-based hypergraph inputs, $f2$ (skipping visited hyperedges) heuristic is one of the most impactful heuristics. With the LiveJournal input, $f1$ (degree) heuristic based algorithm ran out of memory. Hence this data point is missing in the figure. Compared to the other inputs, hyperedges in Livejournal have higher maximum degree-count~(\Cref{tab:input_hypergraph_prop}). Hence, compared to other same-scale hypergraphs, more set-intersections need to be performed when applying only degree-based heuristics. With Friendster and LiveJournal inputs, naive algorithm either did not finish in reasonable time or ran out-of-memory.


\subsubsection{Ablation Study: Effect of different heuristics on different datasets}
We consider the effect of different heuristics on the $s$-overlap computation of  hypergraphs that are drawn from different domains. In addition, we also vary the value of $s$ to see which (set of) heuristic(s) is the most impactful in improving performance. \Cref{fig:effect_of_heuristics} shows the result of our experiments with IMDB, dns-128, and Friendster datasets with $s=2,4,6,8$. Here the speedup is reported by normalizing w.r.t the slowest runtime of all different variations. 

With the IMDB dataset, performance of the algorithms varies significantly, based on the $s$ values. When computing $s$-overlaps with cyber datasets, dns-128, degree-based pruning, $f1$, significantly impacts performance. As we increase the value of $s$ from 4 to 8, the effect of this heuristic remains impactful. 
This implies that pruning the hyperedges based on their sizes eliminated a lot of redundant set intersections. Other heuristics also positively effect the execution time, however less so compared to the  degree-based pruning. With $s=2$ and dns-128 dataset, only our algorithm with all heuristics included ($f0$) ran successfully. Other variations of the algorithm either did not finish in reasonable time or ran out-of-memory. 

On the other hand, with Friendster social network dataset, we observe that 
$f2$ heuristic, i.e. skipping already visited hyperedges for set intersection is the most impactful heuristic. 


Other datasets such as with Web dataset and various $s$ values, only $f0$ finished in a reasonable time limit. Other heuristics are not very helpful in isolation for this dataset. 

\subsubsection{Strong scaling Results}
We also conducted strong scaling experiments with hypergraph inputs from different domains, both with blocked partitioning and cyclic partitioning strategies. Here we increase the number of threads while keeping the dataset (input size) constant. We ran our efficient algorithm ($f0$) with all the heuristics turned on. We report the experimental results with cyclic distribution in \Cref{fig:strong_scaling_cyclic}. 
Here we consider  dns-128 (representing cyber dataset), Friendster (representing social network), IMDB (representing collaboration network), and Web (webgraph) datasets as inputs. As can be seen from the figure, as we increase the number of threads, the performance of the algorithm improves significantly. Blocked distribution demonstrates similar strong scaling behavior for most datasets, 
except for dns-128 and Web. 
With the dns-128 dataset, even with smaller number of threads, cyclic workload successfully distributes workload evenly among the threads to achieve better performance. 
With large-scale hypergraphs such as Web dataset, for which the maximum degree of a hyperedge is 11M, $f0$ with cyclic partitioning achieves ~3x speedup over $f0$ with blocked distribution (the absolute execution time $\approx 8000$ seconds for cyclic vs $\approx 21000$ seconds for blocked range).

\subsubsection{Weak Scaling Results}\label{subsec:weak_scaling}
\begin{wrapfigure}[16]{l}{.2\textwidth}
\vspace{-1em}
  \includegraphics[width=\linewidth]{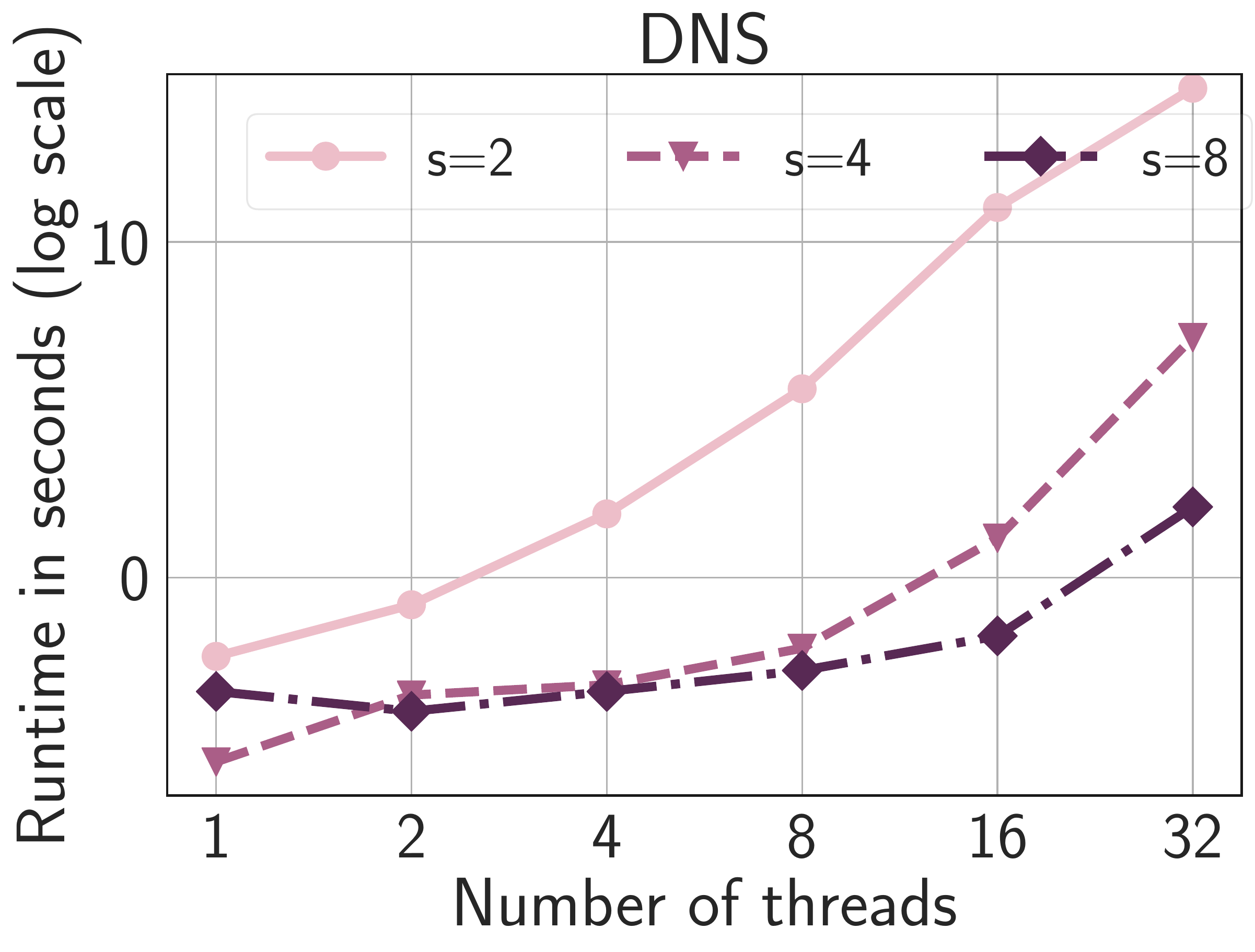}
  \label{fig:dns_full_cyclic}
 \vspace{-1.0em}
 \caption{\small \textit{ Weak scaling results of \Cref{algo:efficient_s_overlap_serial_heuristics} ($f0$) with cyclic distribution with the activeDNS dataset. Here we approximately double the size of the hypergraph (in terms of number of hyperedges) as we increase the number of threads. For example, hypergraph constructed with dns-004 ran with 1 thread, dns-008 with 2 threads, and so on 
.} }
  \label{fig:weak_scaling_hypergraph}
 \vspace{-2.8ex}
\end{wrapfigure}
For weak scaling experiments, we approximately double the size of the hypergraph (workload) as we double the number of threads (computing resources).
We performed weak scaling experiments of \Cref{algo:efficient_s_overlap_serial_heuristics} ($f0$) with the activeDNS dataset and considering cyclic workload distribution strategy. Blocked partition demonstrates similar trend. 
To construct the hypergraphs, we start with 4 input files worth of data (dns-004) and progressively increase the hypergraph size by adding data from more files, as we increase the resources (dns-004 with 1 thread, dns-008 with 2 threads, dns-016 with 4 threads, and so on, up to dns-128).  

We report our weak scaling results in~\Cref{fig:weak_scaling_hypergraph}. As can be seen from the figure, with larger $s$ values, $f0$ with cyclic distribution exhibit better scaling 
(the execution time remains almost constant as we increase the problem size and proportionately increase the computing resources). 
For higher $s$ values 
, the heuristics are more effective to reduce the amount of redundant work.

\subsubsection{Evaluation of relabel-by-degree technique}
We also evaluate how relabeling the hyperedge IDs based on degrees influence the execution time. We have not observed any performance benefit of relabeling when it is considered with blocked partitioning. However, when relabeling is considered in conjunction with cyclic distribution, the performance of the $s$-overlap computation improves. For smaller hypergraphs, the pre-processing time for relabeling can overshadow the benefit of faster $s$-overlap execution time. For example, Friendster takes $\approx 8$ seconds to relabel the hyperedge IDs based on degrees and $\approx 2.7$ seconds to compute $f0$ with cyclic distribution ($s=8$). Without relabeling, the $s-$ overlap computation takes $\approx 3$ seconds. In this case, the overhead is quite significant.

The most interesting application of relabel-by-degree are for computing $s$-overlap with hypergraphs that are larger in size and have a small set of extremely high degree hyperedges. One such example is the Web dataset. For computing the $s$-overlap with $s=11$, $f0$ with only cyclic distribution takes 4230 seconds. On the other hand, when relabel-by-degree is combined with cyclic workload distribution, it takes $\approx 3397$ seconds to execute the $s$-overlap computation step and $\approx 27$ seconds for the pre-processing step, for a total of $\approx 3424$ seconds. In this case, we obtain a speedup of 1.23$\times$ over the version with only cyclic distribution. We also observe that relabel-by-degree in ascending order is considerably faster due to higher cache utility rate according to the profiling done with Intel Vtune profiler. Considering the upper triangular part of the hyperedge adjacency in conjunction with relabel-by-degree in ascending order makes our algorithm cache-friendly.


\begin{wrapfigure}{l}{.35\textwidth}
 \vspace{-2.0ex} 
  \includegraphics[width=\linewidth]{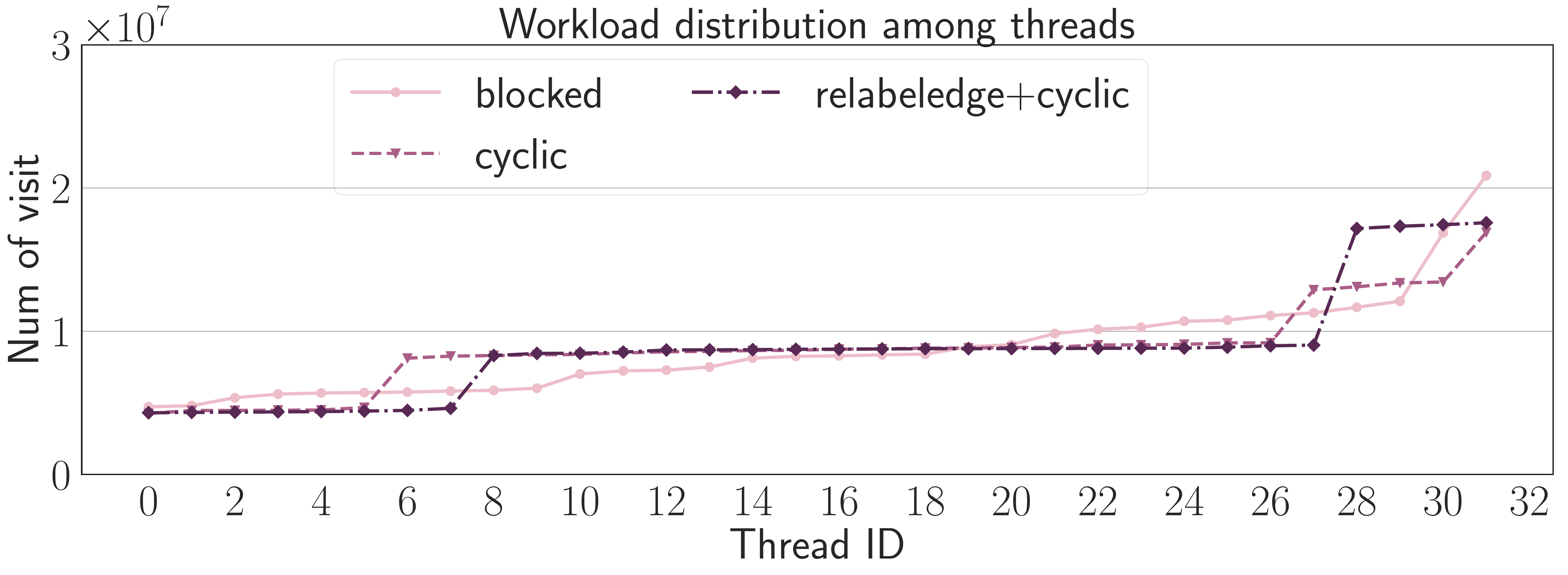}
    \caption{\small \textit{ Workload distribution of Friendster among 32 threads with different partitioning strategies.} }
  \label{fig:workload-dist}
 \vspace{-3.2ex}
\end{wrapfigure}
\subsubsection{Analysis of workload distribution strategies}
To visually assess the effectiveness of different workload distribution techniques among the threads when considering the cyclic and blocked partitioning strategies, we count the total number of hyperedges considered (number of visits) in the innermost loop of our efficient algorithm by each thread. We collect these statistics for the Friendster dataset and report these workloads in~\Cref{fig:workload-dist}. As can be observed from the figure, with cyclic distribution, most of the threads have similar workload profile (the longest running horizontal lines). In contrast, with blocked distribution, the workload varies among the threads. 




\begin{figure*}[tp]
  \centering
  \begin{subfigure}[b]{0.31\textwidth}
    \centering
  \includegraphics[width=\linewidth]{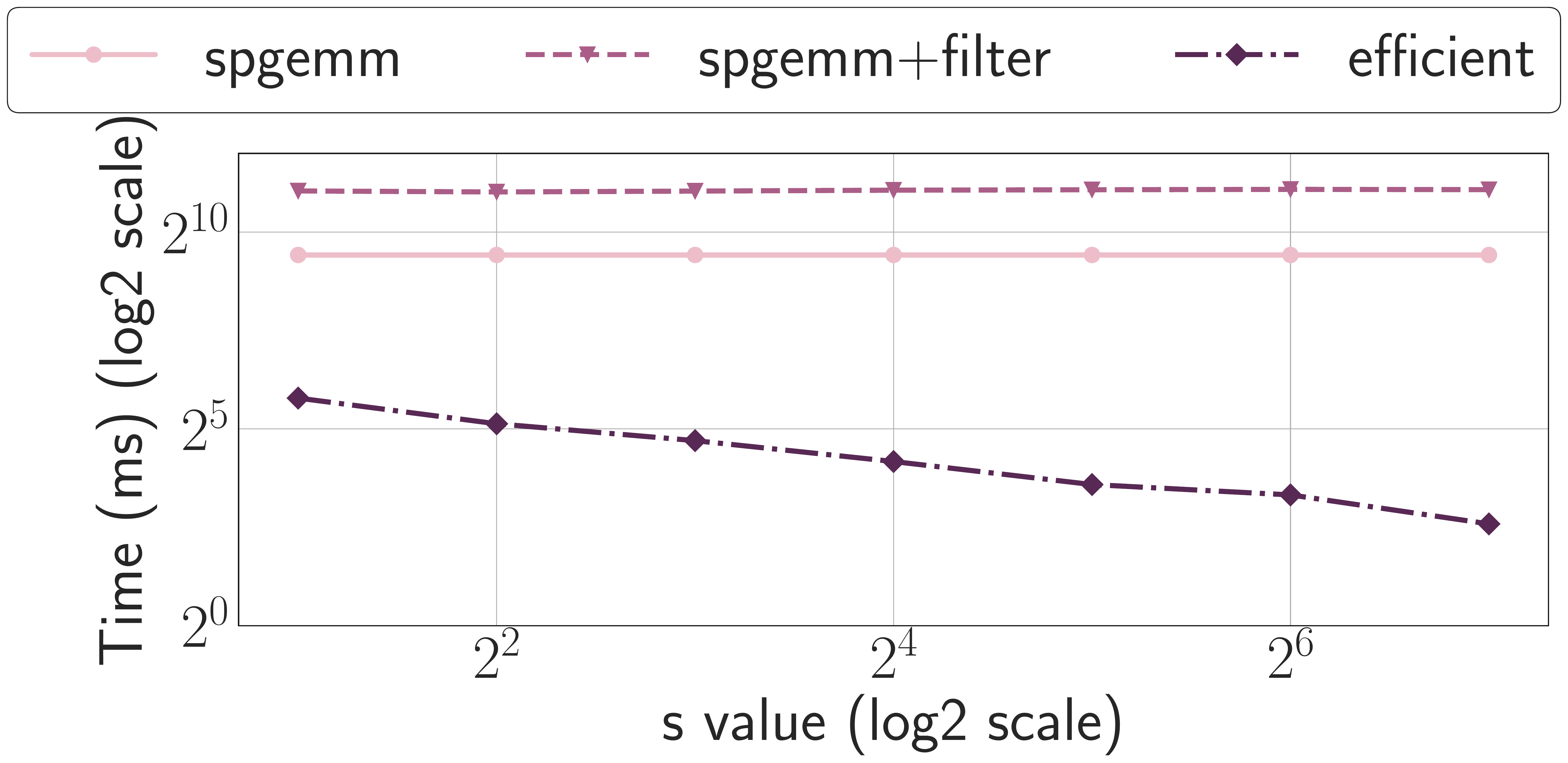}
  \caption{email-EuAll.}
  \label{fig:email-euall-spgemm}
  \end{subfigure} 
 \begin{subfigure}[b]{0.31\textwidth}
    \centering
  \includegraphics[width=\linewidth]{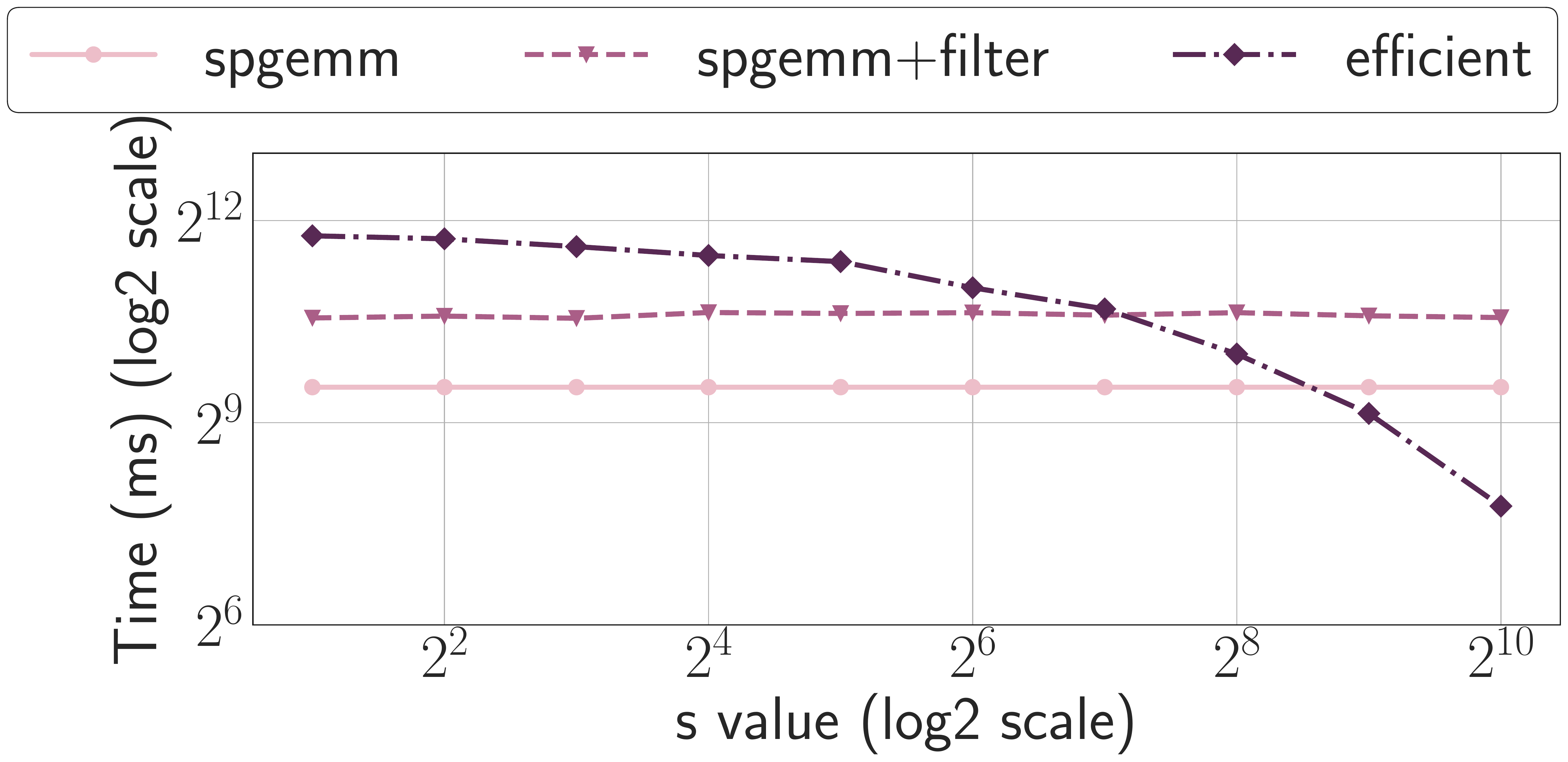}
  \caption{Friendster.}
  \label{fig:friendster_spgemm}
  \end{subfigure} 
  \begin{subfigure}[b]{0.31\textwidth}
    \centering
  \includegraphics[width=\linewidth]{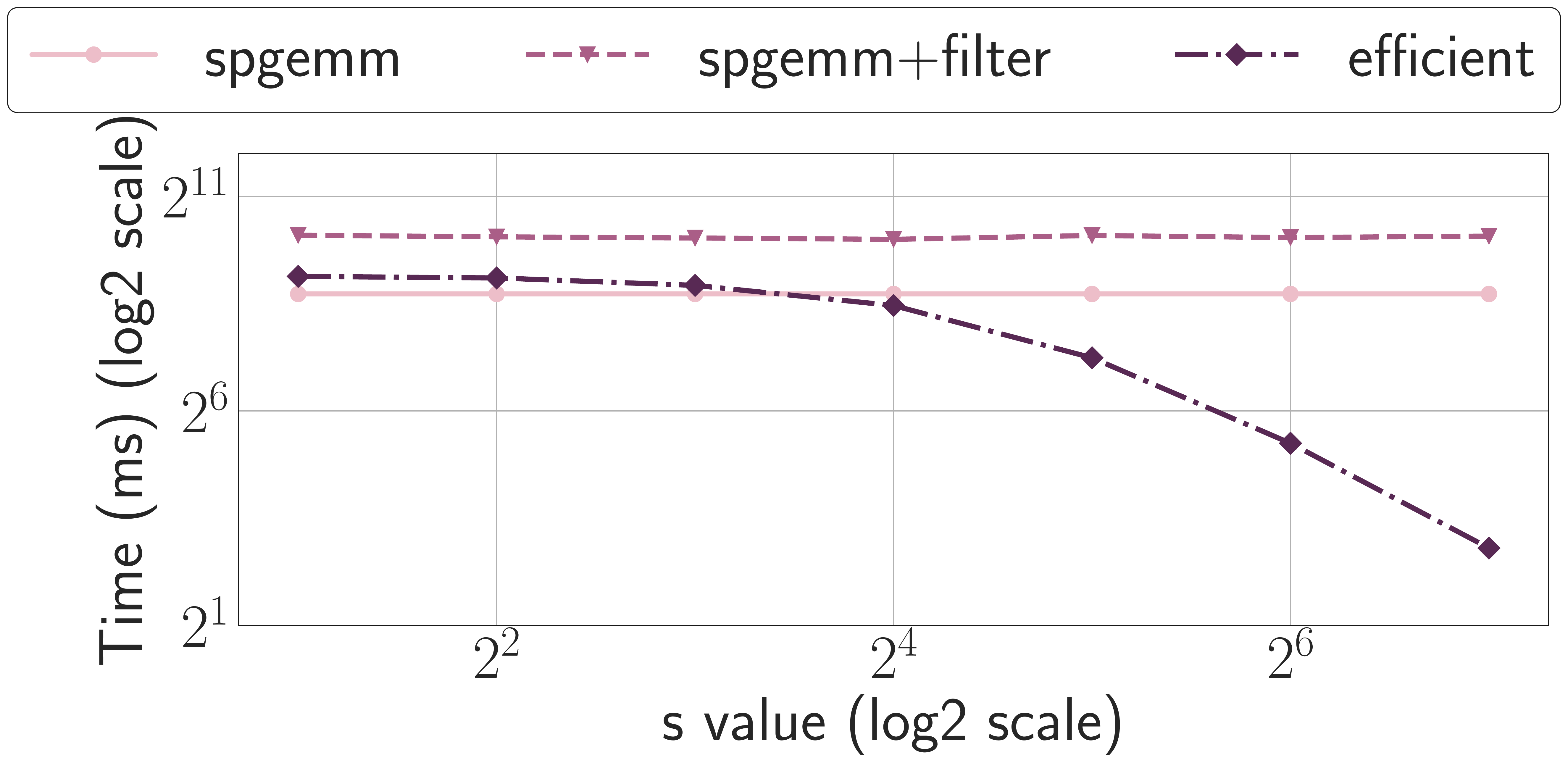}
  \caption{IMDB.}
  \label{fig:imdb_spgemm}
    \end{subfigure} 
  \caption{\small \color{black} \textit{ Comparison of SpGEMM-based approach with \Cref{algo:efficient_s_overlap_serial_heuristics} applying cyclic partitioning, and relabeling by degree (ascending), $f0$} }
  \label{fig:spgemm} 
    \vspace{-1.5em}
\end{figure*}

{\color{black}
\subsubsection{Comparison with SpGEMM-based Approach}
We also benchmark the performance of our $s$-line graph computation algorithm against a state-of-the-art SpGEMM-based library~\cite{nagasaka2019performance,SpGEMMrepo}. We added the filtration step to their code. 
The SpGEMM library first computes $HH^T$ and then a filter is applied to extract the edgelist which satisfies the $s$-overlap constraints. The results are reported in~\Cref{fig:spgemm}. We report the execution time for the SpGEMM step and both SpGEMM plus filtration step separately. With email-EuAll and IMDB datasets, our algorithm is faster with different $s$ values. With the Friendster dataset, computing the $s$ line graph with smaller $s$ value is faster with the SpGEMM library. However, as we increase the $s$ value, our algorithm runs faster than the SpGEMM-based approach. The improvement can be attributed to the degree-based pruning and skipping already visited edges heuristics. Computation of $s$ line graphs with higher $s$ values (for example, $s=1024$) is still relevant for Friendster because, even with such higher $s$ overlap requirement, we found 20 connected components in the constructed $s$ line graph. ``Friends of friends'' are mostly part of the same communities, hence $f2$ heuristic helps improving the performance of our algorithm over off-the-shelf SpGEMM  by pruning redundant work. 
}
\section{Related Work}\label{sec:related_work}
While graph theory dominates network science modeling, hypergraph analytic has also been successfully applied to many real-world datasets and applications, and the value of hypergraphs and related structures are increasingly being recognized, e.g.\ \cite{battiston2020networks,LaNReJ20,joslyn2020hypernetwork,joslyn2020hypergraph}. 

The seminal books~\cite{berge1984hypergraphs,berge1973graphs}, written by Berge, are the firsts of their kinds to provide methodological treatment of hypergraphs as a family of sets and their associated properties. Since then, a plethora of theoretical work has been published  in the context of hypergraphs~\cite{chung1993laplacian,cooper2012spectra,katona1975extremal,dorfler1980category}, and line (intersection) graphs~\cite{bermond1977line,naik2018intersection}. For example, Bermond et al.\cite{bermond1977line} studied $s$-line graph of a hypergraph in details. They show that for any integer $s$ and a graph $G$, there exists a partial hypergraph $H$ of some complete $h$-partite hypergraph such that $G$ is the $s$-line graph of $H$. A recent survey by Niak~\cite{naik2018intersection} on intersection graphs reports recent theoretical developments. 


Only recently, Aksoy et al.~\cite{aksoy2020hypernetwork} proposed a high-order hypergraph walk framework based on $s$-overlaps with \emph{non-uniform hypergraphs}. The paper extends several notions of graph-based techniques to hypergraphs. Although the paper discussed in detail the higher-order analogs of hypergraph methods, the paper does not look into efficient computation of $s$-overlaps and larger dataset, which is the main objective of this paper. Previously, in the context of \emph{uniform hypergraph}, Cooley et al. ~\cite{cooley_evolution_2015} studied the evolution of $s$-connected components in the $k$-uniform binomial random hypergraph. In contrast, we consider more general cases of hypergraphs: both uniform and non-uniform. Thakur and Tripathi ~\cite{thakur_linear_2009} studied the linear connectivity problems in directed hyperpaths, where the hyperconnection between vertices are called $L$-hyperpaths. 

Shun presented a collection of efficient parallel algorithms for hypergraphs in the Hygra framework~\cite{shun_practical_2020}. However, there is no step involved in computing the $s$-line graph of a hypergraph and the algorithms are designed to run directly on the hypergraphs. 
Recent works ~\cite{lumsdaine_2020_triangle, azadevaluation} presented a fast triangle counting implementation with cyclic distribution. As reported in~\cite{azadevaluation}, this is currently the best performing implementation compared to all other state-of-the-art graph frameworks. Our cyclic partitioning strategy is inspired by their work for workload balancing based on cyclic distribution.

Battiston et al.~\cite{battiston2020networks} have provided a comprehensive survey of the state-of-the-art on the structure and dynamics of complex networks beyond dyadic interactions. They discussed structure of systems with higher-order interactions, measures and properties of these systems, random models used for statistical inference in these systems, and dynamics of the systems with higher-order interactions.

\section{Conclusion}\label{sec:conclusion}
In this work, we have proposed efficient, parallel algorithm and heuristics for computing $s$-overlaps and $s$-line graphs. {\color{black} Our algorithm is orders of magnitude (more than $10\times$) faster than the naive algorithm in all cases and the SpGEMM algorithm with filtration in most cases (especially with large $s$ value) for computing $s$-overlap.} To balance workload among threads, we consider different workload partitioning techniques including blocked range and we implement and apply a customized cyclic range to achieve better performance. Computing $s$-overlaps is the fundamental step for creating line graphs. Once we compute the $s$-line graphs, many important features of the original hypergraph are retained in the $s$-line graph and yet we can leverage highly tuned graph algorithms to analyze the data. In this way, analysts can leverage the multi-way relationships in hypergraph to find meaningful relationships.


\section{}\label{sec:ack}
This work was partially funded under the High Performance Data Analytics
(HPDA) program at the Department of Energy's Pacific Northwest National
Laboratory, and alos partially supported by NSF SI2-SSE Award 1716828 and NSF award IIS-1553528. PNNL Information Release: PNNL-SA-164086. Pacific Northwest
National Laboratory is operated by Battelle Memorial Institute under Contract DE-ACO6-76RL01830.

\bibliographystyle{IEEEtranS}
\bibliography{main}
\end{document}